\documentclass[aps,prd,twocolumn,nofootinbib,,superscriptaddress,showkeys]{revtex4}
\usepackage{amssymb,amsmath,graphicx,enumerate}
\usepackage{subfigure}
\usepackage{dcolumn,booktabs}
\usepackage[dvipdfm,colorlinks=true,citecolor=blue,pdfstartview=FitH]{hyperref}

\usepackage{color,framed} %making color table background.
\definecolor{shadecolor}{rgb}{0,0,1} %making color table background.  \begin{shaded} ****\end{shaded}.

\def\bea{\begin{equation}}
\def\eea{\end{equation}}
\newcommand{\hbts}{hidden bottom and charm tetraquarks}

\newcommand{\twh}{tetraquarks with hidden bottom and charm}
\newcommand{\twhf}{$(Qq)(\bar{Q}\bar{q}')$}
\newcommand{\rt}{Regge trajectory}
\newcommand{\rts}{Regge trajectories}

\newcommand{\trs}{trajectories}
\newcommand{\bfr}{{\bf r}}
\newcommand{\bfp}{{\bf p}}
\newcommand{\bfpa}{{|\bf p|}}
\newcommand{\gev}{{\rm GeV}}

\newcommand{\cltb}{$\bar{3}_c$}
\newcommand{\cltba}{\bar{3}_c}
\newcommand{\cls}{$6_c$}
\newcommand{\dqs}{$(qq')$}
\newcommand{\bqm}{\,{\pmb{?}}}

\begin{document}
\title{$\lambda$ and $\rho$ Regge trajectories for hidden bottom and charm tetraquarks $(Qq)(\bar{Q}\bar{q}')$}
\author{Jia-Qi Xie}
\email{1462718751@qq.com}
\affiliation{School of Physics and Information Engineering, Shanxi Normal University, Taiyuan 030031, China}
\author{He Song}
\email{songhe\_22@163.com}
\affiliation{School of Physics and Information Engineering, Shanxi Normal University, Taiyuan 030031, China}
\author{Xia Feng}
\email{sxsdwxfx@163.com}
\affiliation{School of Physics and Information Engineering, Shanxi Normal University, Taiyuan 030031, China}
\author{Jiao-Kai Chen}
\email{chenjk@sxnu.edu.cn, chenjkphy@outlook.com (corresponding author)}
\affiliation{School of Physics and Information Engineering, Shanxi Normal University, Taiyuan 030031, China}
%\date{\today}

\begin{abstract}
We propose the Regge trajectory relations for the heavy tetraquarks $(Qq)(\bar{Q}\bar{q}')$ $(Q=b,\,c;\,q,\,q'=u,\,d,\,s)$ with hidden bottom and charm. By employing the new relations, both the $\lambda$-trajectories and the $\rho$-trajectories for the tetraquarks $(Qq)(\bar{Q}\bar{q}')$ can be discussed. The masses of the $\lambda$-mode excited states and the $\rho$-mode excited states are estimated, and they agree with other theoretical predictions.
We show that the behaviors of the $\rho$-trajectories are different from those of the $\lambda$-trajectories. The $\rho$-trajectories behave as $M{\sim}x_{\rho}^{1/2}$ $(x_{\rho}=n_r,\,l)$ while the $\lambda$-trajectories behave as $M{\sim}x_{\lambda}^{2/3}$ $(x_{\lambda}=N_r,\,L)$.
Moreover, the Regge trajectory behaviors for other types of tetraquarks are investigated based on the spinless Salpeter equation. We show that both the $\lambda$-trajectories and the $\rho$-trajectories are concave downward in the $(M^2,\,x)$ plane. The Regge trajectories for the tetraquarks containing the light diquark and/or the light antidiquark also are concave in the $(M^2,\,x)$ plane when the masses of the light constituents are included and the confining potential is linear.
\end{abstract}

\maketitle

%%%%%%%%%%%%%%%%%%%%%%%%%%%%%%%%%%%%%%%%%%%%%%%%%%%%%%%%%%%%%%%%%%%%%%%%%%

\section{Introduction}
As a type of exotic hadrons, tetraquarks will enhance the study of hadrons and will provide new probes for the understanding of QCD \cite{Jaffe:1976ig,Jaffe:1976ih,pdg:2024,Gross:2022hyw,Bicudo:2022cqi,Liu:2019zoy,
Jaffe:2004ph,Esposito:2016noz}. The heavy tetraquarks $(Qq)(\bar{Q}\bar{q}')$ $(Q=b,\,c;\, q,q'=u,\,d,\,s)$ with hidden bottom and charm have been studied through various methods. These approaches include the relativistic quark model based on the quasipotential approach in QCD \cite{Ebert:2008se,Faustov:2021hjs,Ebert:2008kb},  the non-relativistic potential model \cite{Tiwari:2022azj,Lundhammar:2020xvw}, the spinless Salpeter equation \cite{Ferretti:2020ewe,Giannuzzi:2023vrx}, the relativistic four-body Faddeev-Yakubovsky equation \cite{Hoffer:2024alv}, the dynamical diquark model \cite{Giron:2020qpb,Lebed:2024zrp,Giron:2020fvd,Giron:2021sla}, the homogeneous Lippmann-Schwinger integral equation \cite{Hadizadeh:2015cvx}, the effective Hamiltonian in the adopted chromomagnetic interaction model \cite{Wu:2018xdi}, the chiral quark model \cite{Liu:2022vyy}, the sum rules \cite{Chen:2017dpy,Wang:2019mxn}, the relativistic four-quark equations in the framework of coupled-channel
formalism \cite{Gerasyuta:2008cb}, the constituent quark model \cite{Maiani:2004vq}, among others.

The {\rt}\footnote{A {\rt} of hadrons is assumed to be written as $M=m_R+\beta_x(x+c_0)^{\nu}$ $(x=l,\,n_r)$, where $M$ is mass of the bound state, $l$ is the orbital angular momentum, and $n_r$ is the radial quantum number. $m_R$ and $\beta_x$ are parameters. For simplicity, the plots in the $(M,\,x)$ plane \cite{Chen:2022flh,Chen:2023cws}, in the $(M,\,(x-c_0)^{\nu})$ plane \cite{Burns:2010qq}, in the $(M^2,\,x)$ plane \cite{Chen:2018nnr,Chen:2021kfw}, in the $((M-m_R)^2,\,x)$ plane \cite{Chen:2023djq,Chen:2023web} or in the $((M-m_R)^{1/{\nu}},\,x)$ plane [see Fig. \ref{fig:rgapp} in this work] are all called the Chew-Frautschi plots. The {\rts} can be plotted in these different planes. } is one of the effective approaches widely used in the study of hadron spectra \cite{Regge:1959mz,Chew:1961ev,Chew:1962eu,Nambu:1978bd,
Collins:1971ff,Inopin:1999nf,Irving:1977ea}.
Ref. \cite{Selem:2006nd} gives a compact and predictive unified picture of mesons, baryons, and tetraquarks with the loaded flux-tube model. Ref. \cite{Nielsen:2018uyn} presents relations between the Regge trajectories of mesons, baryons, and tetraquarks based on the Holographic light front QCD. In Ref. \cite{Sonnenschein:2018fph}, the masses of mesons, baryons, glueballs, and tetraquarks are predicted by the {\rts} obtained from the holography inspired stringy hadron model. In Ref. \cite{MartinContreras:2020cyg}, mesons, hybrid mesons and multiquark states are discussed by the nonlinear {\rts} in the model AdS/QCD with dilaton. In Refs. \cite{Chen:2023web,Chen:2023djq}, we discuss the universal descriptions of the heavy-heavy and heavy-light systems by using the {\rts}.
In preceding works, the Regge trajectories for the tetraquarks are the $\lambda$-trajectories. To our knowledge, no studies address the $\rho$-trajectories for tetraquarks although there are many studies on the $\rho$-excited states of tetraquarks.
In the diquark picture, the hidden bottom tetraquarks are composed of one heavy-light diquark $(Qq)$ and one heavy-light antidiquark $(\bar{Q}\bar{q}')$. By applying the {\rt} relation for the heavy-light diquarks \cite{Chen:2023cws} and the {\rt} formula for the heavy-heavy systems \cite{Chen:2023djq}, we present the {\rt} relations for the {\hbts}. The masses of the $\lambda$-mode excited and $\rho$-mode excited {\twh} are estimated. Both the $\lambda$-trajectories and $\rho$-trajectories are investigated.

The paper is organized as follows: In Sec. \ref{sec:rgr}, the {\rts} for the {\hbts} are discussed. The conclusions are presented in Sec. \ref{sec:conc}. The discussions on the {\rt} behaviors for various tetraquarks are in the appendix \ref{app:c}.

\section{{\rts} for the {\twh}}\label{sec:rgr}
In this section, with the help of the diquark {\rts} \cite{Chen:2023cws,Feng:2023txx,Chen:2023ngj}, we investigate the {\rt} behaviors for various tetraquarks. The {\rt} relations are presented, which can be employed to discuss both the $\lambda$-trajectories and the $\rho$-trajectories.

\subsection{Preliminary}\label{subsec:prelim}

\begin{figure}[!phtb]
\centering
\includegraphics[width=0.25\textheight]{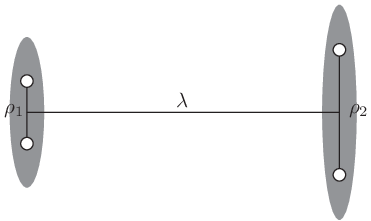}
\caption{Schematic diagram of the tetraquarks in the diquark-antidiquark picture.}\label{fig:tr}
\end{figure}

In the diquark picture, tetraquarks consist of one diquark and one antidiquark, see Fig. \ref{fig:tr}. $\rho_1$ ($\rho_2$) separates the quarks (antiquarks) in the diquark (antidiquark), and $\lambda$ separates the diquark and the antidiquark. There exist three excited modes: the $\rho_1$-mode involves the radial and orbital excitation in the diquark, the $\rho_2$-mode involves the radial and orbital excitation in the antidiquark, and the $\lambda-$mode involves the radial or orbital excitation between the diquark and antidiquark. Consequently, there exist three series of {\rts}: two series of $\rho$-{\rts} and one series of $\lambda$-{\rts}.

In $SU_c(3)$, there is attraction between quark pairs $(qq')$ in the color antitriplet channel and this is just twice weaker than in the color singlet $q\bar{q}'$ in the one-gluon exchange approximation.
Only the {\cltb} representation of $SU_c(3)$ is considered in the present work; the {\cls} representation is not considered. The tetraquarks considered are composed of a diquark and an antidiquark in color $\bar{3}$ and $3$ configuration.

In the diquark picture, the state of tetraquark is denoted as
\bea\label{tetnot}
\left((q_1q'_1)^{{\bar{3}_c}}_{n_1^{2s_1+1}l_{1j_1}}
(\bar{q}_2\bar{q}'_2)^{{{3}_c}}_{n_2^{2s_2+1}l_{2j_2}}\right)^{1_c}_{N^{2j+1}L_J},
\eea
where {\cltb} denotes the color antitriplet state of diquark, and $1_c$ represents the color singlet state of tetraquark. [The superscript $1_c$ is often omitted because the tetraquarks are colorless.] The diquark $(q_1q'_1)$ is $\{q_1q'_1\}$ or $[q_1q'_1]$. The antidiquark $(\bar{q}_2\bar{q}'_2)$ is $\{\bar{q}_2\bar{q}'_2\}$ or $[\bar{q}_2\bar{q}'_2]$. $\{qq'\}$ and $[qq']$ indicate the permutation symmetric and antisymmetric flavor wave functions, respectively. $N=N_{r}+1$, where $N_{r}=0,\,1,\,\cdots$. $n_{1,2}=n_{r_{1,2}}+1$, where $n_{r_{1,2}}=0,\,1,\,\cdots$. $N_r$, $n_{r_1}$ and $n_{r_2}$ are the radial quantum numbers of the tetraquark, diquark 1, antidiquark 2, respectively. The completely antisymmetric states for the diquarks in {\cltb} are listed in Table \ref{tab:dqstates}.
$\vec{J}=\vec{L}+\vec{j}$, $\vec{j}=\vec{j}_1+\vec{j}_2$, $\vec{j}_1=\vec{s}_1+\vec{l}_1$, $\vec{j}_2=\vec{s}_2+\vec{l}_2$.
$\vec{J}$, $\vec{j}_1$ and $\vec{j}_2$ are the spins of tetraquark, diquark 1, antidiquark 2, respectively. $\vec{j}$ is the summed spin of diquarks and antidiquarks in the tetraquark. $L$, $l_{1}$ and $l_2$ are the orbital quantum numbers of tetraquark, diquark 1 and antidiquark 2, respectively. $\vec{s}_{1}$ ($\vec{s}_{2}$) is the summed spin of quarks (antiquarks) in the diquark (antidiquark).

\subsection{{\rt} relations for the {\hbts} }
The {\twh} consists of one heavy-light diquark $(Qq)$ and one heavy-light antidiquark $(\bar{Q}\bar{q}')$. Both $(Qq)$ and $(\bar{Q}\bar{q}')$ are heavy because the bottom/charm quark are heavy. Therefore, the tetraquarks {\twhf} are the heavy-heavy systems in case of the $\lambda$-mode excitations. Meanwhile, the {\rts} for {\twhf} exhibit the properties of the heavy-light systems in case of the $\rho$-mode excitations. According to Eqs. (\ref{t2q}) and (\ref{pa2qQ}), we list the relations for the {\twh}
\begin{align}\label{t2qx}
M&=m_{R{\lambda}}+\beta_{x_{\lambda}}(x_{\lambda}+c_{0x_{\lambda}})^{2/3}\;(x_{\lambda}=L,\,N_r),\nonumber\\
M_{\rho_1}&=m_{R\rho_1}+\beta_{x_{\rho_1}}\sqrt{x_{\rho_1}+c_{0x_{\rho_1}}}\;(x_{\rho_1}=l_{1},\,n_{r_1}),\nonumber\\
M_{\rho_2}&=m_{R\rho_2}+\beta_{x_{\rho_2}}\sqrt{x_{\rho_2}+c_{0x_{\rho_2}}}\;(x_{\rho_2}=l_{2},\,n_{r_2}),
\end{align}
where
\begin{align}\label{pa2qQx}
m_{R{\lambda}}&=M_{\rho_1}+M_{\rho_2}+C,\nonumber\\
m_{R\rho_1}&=m_{Q}+m_{q}+C/2,\nonumber\\
m_{R\rho_2}&=m_{Q}+m_{q^{\prime}}+C/2,\nonumber\\
\beta_{L}&=\frac{3}{2}\left(\frac{\sigma^2}{\mu_{\lambda}}\right)^{1/3}c_{fL},\nonumber\\ \beta_{N_r}&=\frac{(3\pi)^{2/3}}{2}\left(\frac{\sigma^2}{\mu_{\lambda}}\right)^{1/3}c_{fN_r},\nonumber\\
\mu_{\lambda}&=\frac{M_{\rho_1}M_{\rho_2}}{M_{\rho_1}+M_{\rho_2}},\nonumber\\
\beta_{l_1}&=\sqrt{2\sigma}c_{fl_1},\quad \beta_{n_{r_1}}=\sqrt{\pi\sigma}c_{fn_{r_1}},\nonumber\\
\beta_{l_2}&=\sqrt{2\sigma}c_{fl_2},\quad \beta_{n_{r_2}}=\sqrt{\pi\sigma}c_{fn_{r_2}}.
\end{align}
$M$, $M_{\rho_1}$, $M_{\rho_2}$, $m_{Q}$, and $m_{q}$ are the tetraquark mass, the diquark mass, the antidiquark mass, the heavy quark mass, and the light quark mass, respectively. $\sigma$ is the string tension. $C$ is a fundamental parameter. $c_{fx}$ are theoretically equal to one but are fitted in practice. $c_{0x}$ vary with {\rts}.

\subsection{Parameters}
The quark masses, the string tension $\sigma$ and the parameter $C$ are from Ref. \cite{Faustov:2021hjs}. The parameters for the heavy-light diquarks $(bu)$, $(bs)$, $(cu)$, and $(cs)$ are from Ref. \cite{Chen:2023cws} and listed in Table \ref{tab:parmv}. With these parameters determined, the $\rho$-modes and the diquark masses can be discussed, see Eqs. (\ref{t2qx}) and (\ref{pa2qQx}).
To discuss the masses of the {\hbts} and the $\lambda$-modes excited states,
the parameters $c_{fL}$, $c_{fN_r}$, $c_{0{L}}$ and $c_{0N_r}$ should be determined, see Eqs. (\ref{fitcfxl}) and (\ref{fitcfxnr}). More discussions are in section \ref{sec:appcfx}.
Once all parameters are determined, the {\rts} can be discussed and can be used to estimate the masses of the {\twh}.

\begin{table}[!phtb]
\caption{The values of parameters \cite{Chen:2023cws,Faustov:2021hjs}.}  \label{tab:parmv}
\centering
\begin{tabular*}{0.47\textwidth}{@{\extracolsep{\fill}}cl@{}}
\hline\hline
          & $m_{u,d}=0.33\; {\gev}$, \; $m_s=0.50\; {\gev}$, \; $m_b=4.88\; {\gev}$  \\
          & $m_c=1.55\; {\gev}$, \; $\sigma=0.18\; {\gev^2}$,\; $C=-0.3\; {\gev}$ \\
$(bu)$    & $c_{fn_{r_1}}=c_{fn_{r_2}}=0.988$,\; $c_{fl_1}=c_{fl_2}=0.965$  \\
          & $c_{0n_{r_1}}(1^1s_0)=c_{0n_{r_2}}(1^1s_0)=0.125$\\
          & $c_{0n_{l_1}}(1^1s_0)=c_{0n_{l_2}}(1^1s_0)=0.18$\\
          & $c_{0n_{r_1}}(1^3s_1)=c_{0n_{r_2}}(1^3s_1)=0.155$\\
          & $c_{0n_{l_1}}(1^3s_1)=c_{0n_{l_2}}(1^3s_1)=0.22$\\
$(bs)$    & $c_{fn_{r_1}}=c_{fn_{r_2}}=0.953$,\; $c_{fl_1}=c_{fl_2}=0.919$  \\
          & $c_{0n_{r_1}}(1^1s_0)=c_{0n_{r_2}}(1^1s_0)=0.08$\\
          & $c_{0n_{l_1}}(1^1s_0)=c_{0n_{l_2}}(1^1s_0)=0.115$\\
          & $c_{0n_{r_1}}(1^3s_1)=c_{0n_{r_2}}(1^3s_1)=0.11$\\
          & $c_{0n_{l_1}}(1^3s_1)=c_{0n_{l_2}}(1^3s_1)=0.16$\\
$(cu)$    & $c_{fn_{r_1}}=c_{fn_{r_2}}=1.000$,\; $c_{fl_1}=c_{fl_2}=1.038$  \\
          & $c_{0n_{r_1}}(1^1s_0)=c_{0n_{r_2}}(1^1s_0)=0.065$\\
          & $c_{0n_{l_1}}(1^1s_0)=c_{0n_{l_2}}(1^1s_0)=0.095$\\
          & $c_{0n_{r_1}}(1^3s_1)=c_{0n_{r_2}}(1^3s_1)=0.17$\\
          & $c_{0n_{l_1}}(1^3s_1)=c_{0n_{l_2}}(1^3s_1)=0.19$\\
$(cs)$    & $c_{fn_{r_1}}=c_{fn_{r_2}}=1.016$,\; $c_{fl_1}=c_{fl_2}=1.015$  \\
          & $c_{0n_{r_1}}(1^1s_0)=c_{0n_{r_2}}(1^1s_0)=0.03$\\
          & $c_{0n_{l_1}}(1^1s_0)=c_{0n_{l_2}}(1^1s_0)=0.055$\\
          & $c_{0n_{r_1}}(1^3s_1)=c_{0n_{r_2}}(1^3s_1)=0.095$\\
          & $c_{0n_{l_1}}(1^3s_1)=c_{0n_{l_2}}(1^3s_1)=0.135$\\
\hline
\hline
\end{tabular*}
\end{table}

\subsection{$\lambda$- and $\rho$-{\trs} for the hidden bottom tetraquarks}\label{subsec:rts}

\begin{table*}[!phtb]
\caption{The masses of the $\lambda$-mode excited states of the hidden bottom tetraquarks (in ${\gev}$). The notation in Eq. (\ref{tetnot}) is rewritten as $|n_1^{2s_1+1}l_{1j_1},n_2^{2s_2+1}l_{2j_2},N^{2j+1}L_J\rangle$. $n=u,\,d$. Eqs. (\ref{t2qx}), (\ref{fitcfxl}) and (\ref{fitcfxnr}) are used.}  \label{tab:masslambda}
\centering
\begin{tabular*}{1.0\textwidth}{@{\extracolsep{\fill}}cccc@{}}
\hline\hline
  $|n_1^{2s_1+1}l_{1j_1},n_2^{2s_2+1}l_{2j_2},N^{2j+1}L_J\rangle$        & $(bn)(\bar{b}\bar{n})$  &  $(bn)(\bar{b}\bar{s})$  &  $(bs)(\bar{b}\bar{s})$ \\
\hline
 $|1^1s_0, 1^1s_0, 1^1S_0\rangle$  & 10.46   & 10.57   &10.68 \\
 $|1^1s_0, 1^1s_0, 2^1S_0\rangle$  & 10.91   & 11.02   &11.12  \\
 $|1^1s_0, 1^1s_0, 3^1S_0\rangle$  & 11.21   & 11.32   &11.43  \\
 $|1^1s_0, 1^1s_0, 4^1S_0\rangle$  & 11.47   & 11.57   &11.68  \\
 $|1^1s_0, 1^1s_0, 5^1S_0\rangle$  & 11.70   & 11.80   &11.91  \\
\hline
 $|1^1s_0, 1^1s_0, 1^1S_0\rangle$  & 10.46   & 10.57   & 10.68 \\
 $|1^1s_0, 1^1s_0, 1^1P_1\rangle$  & 10.78   & 10.89   & 11.00 \\
 $|1^1s_0, 1^1s_0, 1^1D_2\rangle$  & 11.01   & 11.12   & 11.22 \\
 $|1^1s_0, 1^1s_0, 1^1F_3\rangle$  & 11.20   & 11.31   & 11.41 \\
 $|1^1s_0, 1^1s_0, 1^1G_4\rangle$  & 11.37   & 11.48   & 11.59 \\
 $|1^1s_0, 1^1s_0, 1^1H_5\rangle$  & 11.53   & 11.64   & 11.74 \\
\hline\hline
\end{tabular*}
\end{table*}

\begin{figure*}[!phtb]
\centering
\subfigure[]{\label{subfigure:cfa}\includegraphics[scale=0.6]{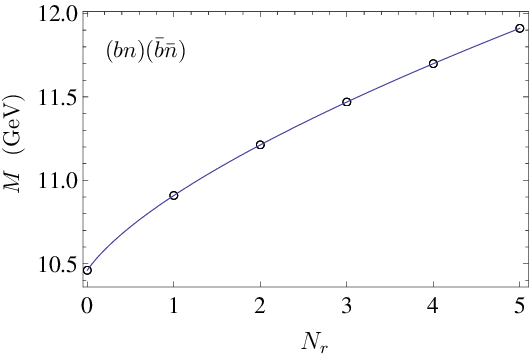}}
\subfigure[]{\label{subfigure:cfa}\includegraphics[scale=0.6]{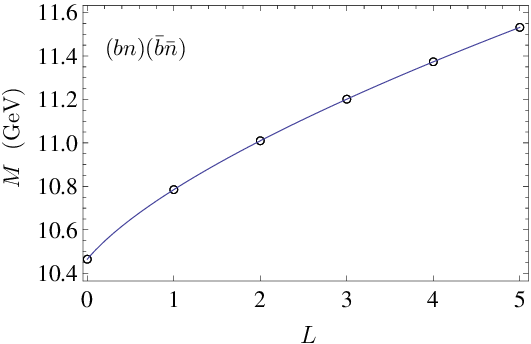}}
\subfigure[]{\label{subfigure:cfa}\includegraphics[scale=0.6]{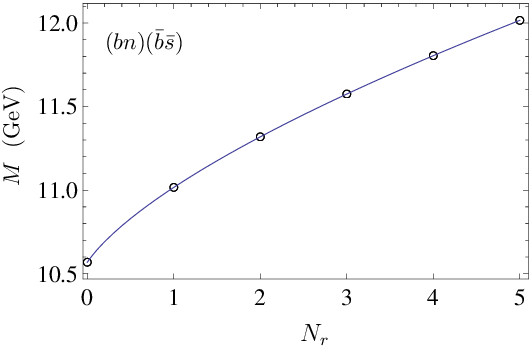}}
\subfigure[]{\label{subfigure:cfa}\includegraphics[scale=0.6]{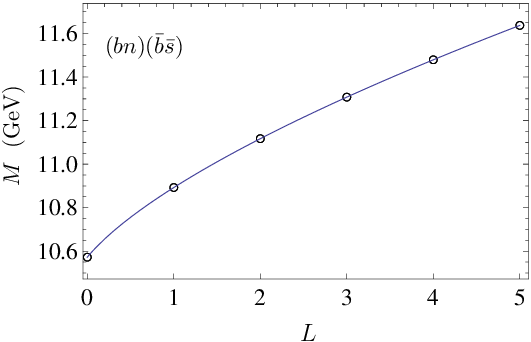}}
\subfigure[]{\label{subfigure:cfa}\includegraphics[scale=0.6]{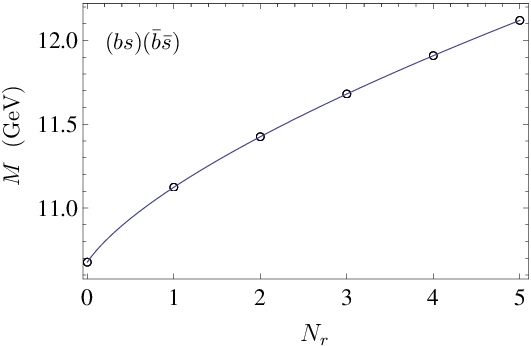}}
\subfigure[]{\label{subfigure:cfa}\includegraphics[scale=0.6]{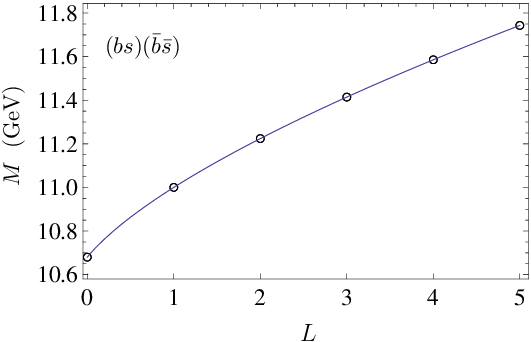}}
\caption{The $\lambda$-trajectories for the hidden bottom tetraquarks. Circles represent the predicted data and the blue lines are the $\lambda$-trajectories, see Eq. (\ref{t2qx}). Data are listed in Table \ref{tab:masslambda}. $M$ is the bound state mass. $N_r$ and $L$ are the radial and orbital quantum numbers for the $\lambda$-mode, respectively.}\label{fig:tlam}
\end{figure*}

\begin{table*}[!phtb]
\caption{Comparison of theoretical predictions for the masses of the hidden bottom tetraquarks (in GeV). The charge conjugation $C$ is defined only for $q=q'$.}  \label{tab:masscomp}
\centering
\begin{tabular*}{1.0\textwidth}{@{\extracolsep{\fill}}ccccccccc@{}}
\hline\hline
 $J^{PC}$ &  $|n_1^{2s_1+1}l_{1j_1},n_2^{2s_2+1}l_{2j_2},N^{2j+1}L_J\rangle$
          & Tetraquark  &   Our  & FGS \cite{Faustov:2021hjs} & FS \cite{Ferretti:2020ewe} & TR \cite{Tiwari:2022azj} & HK \cite{Hadizadeh:2015cvx} \\
\hline
 $0^{++}$ &  $|1^1s_0, 1^1s_0, 1^1S_0\rangle$ &
                $(bn)(\bar{b}\bar{n})$ & 10.46 & 10.471 &       &        &10.410 \\
         &   &  $(bn)(\bar{b}\bar{s})$ & 10.57 &10.572  &10.407 &10.522  & \\
         &   &  $(bs)(\bar{b}\bar{s})$ & 10.68 &10.662  &       &10.711  &10.613\\
%%%%%%%%%%%%%%
         &  $|1^1s_0, 1^1s_0, 2^1S_0\rangle$
             & $(bn)(\bar{b}\bar{n})$  & 10.91 & 10.917 &       &      &10.899\\
         &   &  $(bn)(\bar{b}\bar{s})$ & 11.02 &11.018  &10.909 &11.085&\\
         &   &  $(bs)(\bar{b}\bar{s})$ & 11.12 &11.111  &       &11.244&11.100 \\
%%%%%%%%%%%%%%%%%%
  $1^{--}$  &$|1^1s_0, 1^1s_0, 1^1P_1\rangle$  &
               $(bn)(\bar{b}\bar{n})$ & 10.78 &10.807 &       &       &10.806 \\
        &   &  $(bn)(\bar{b}\bar{s})$ & 10.89 &10.907 &10.790 &11.024 & \\
        &   &  $(bs)(\bar{b}\bar{s})$ & 11.00 &11.002 &       &11.184 &11.009 \\
\hline\hline
\end{tabular*}
\end{table*}

The tetraquarks with hidden bottom include the neutral states $(bu)(\bar{b}\bar{u})$
and $(bd)(\bar{b}\bar{d})$, the charged states $(bu)(\bar{b}\bar{d})$
and $(bd)(\bar{b}\bar{u})$, the states with open strangeness $(bs)(\bar{b}\bar{n})$ and $(bn)(\bar{b}\bar{s})$ ($n=u,\,d$), and the states with hidden strangeness $(bs)(\bar{b}\bar{s})$. We assume $m_u=m_d$, therefore, the masses for diquarks $(bd)$ and $(bu)$ are equal and it holds for the tetraquarks as well.

When calculating the $\lambda$-mode excitations, the scalar diquark and the scalar antidiquark are used. The radially and orbitally excited states of the $\lambda$-mode are calculated by using Eqs. (\ref{t2qx}), (\ref{fitcfxl}) and (\ref{fitcfxnr}) and parameters in Table \ref{tab:parmv}. The calculated results are listed in Table \ref{tab:masslambda}. More results can be easily obtained in a similar manner. The $\lambda$-trajectories for the hidden bottom tetraquarks are shown in Fig. \ref{fig:tlam}.

\begin{table*}[!htbp]
\caption{The masses of the $\rho$-mode excited states of the hidden bottom tetraquarks (in ${\gev}$). The notation in Eq. (\ref{tetnot}) is rewritten as $|n_1^{2s_1+1}l_{1j_1},n_2^{2s_2+1}l_{2j_2},N^{2j+1}L_J\rangle$. $(M^{\prime},\,M^{\prime\prime})$ is for
$\left(|n_1^{2s_1+1}l_{1j_1},n_2^{2s_2+1}l_{2j_2},N^{2j+1}L_J\rangle,\;
|n_2^{2s_2+1}l_{2j_2},n_1^{2s_1+1}l_{1j_1},N^{2j+1}L_J\rangle\right)$. For $(bn)(\bar{b}\bar{n})$ and $(bs)(\bar{b}\bar{s})$, $M'=M^{\prime\prime}$. $n=u,\,d$. Eqs. (\ref{t2qx}), (\ref{fitcfxl}) and (\ref{fitcfxnr}) are used.}  \label{tab:massrho}
\centering
\begin{tabular*}{1.0\textwidth}{@{\extracolsep{\fill}}cccc@{}}
\hline\hline
  $|n_1^{2s_1+1}l_{1j_1},n_2^{2s_2+1}l_{2j_2},N^{2j+1}L_J\rangle$        & $(bn)(\bar{b}\bar{n})$  &  $(bn)(\bar{b}\bar{s})$  &  $(bs)(\bar{b}\bar{s})$ \\
  &        $M^{\prime}=M^{\prime\prime}$         &  $(M^{\prime},\,M^{\prime\prime})$              & $M^{\prime}=M^{\prime\prime}$   \\
\hline
 $|1^1s_0, 1^1s_0, 1^1S_0\rangle$  &10.46  &(10.57,\,10.57)   &10.68 \\
 $|2^1s_0, 1^1s_0, 1^1S_0\rangle$  &10.98  &(11.08,\,11.10)   &11.21 \\
 $|3^1s_0, 1^1s_0, 1^1S_0\rangle$  &11.27  &(11.37,\,11.38)   &11.49 \\
 $|4^1s_0, 1^1s_0, 1^1S_0\rangle$  &11.49  &(11.60,\,11.61)   &11.71 \\
 $|5^1s_0, 1^1s_0, 1^1S_0\rangle$  &11.69  &(11.79,\,11.79)   &11.90 \\
\hline
 $|1^3s_1, 1^1s_0, 1^3S_1\rangle$  &10.49  &(10.60,\,10.60)   &10.71 \\
 $|2^3s_1, 1^1s_0, 1^3S_1\rangle$  &10.99  &(11.09,\,11.11)   &11.22 \\
 $|3^3s_1, 1^1s_0, 1^3S_1\rangle$  &11.27  &(11.38,\,11.39)   &11.50 \\
 $|4^3s_1, 1^1s_0, 1^3S_1\rangle$  &11.50  &(11.61,\,11.61)   &11.72 \\
 $|5^3s_1, 1^1s_0, 1^3S_1\rangle$  &11.69  &(11.80,\,11.80)   &11.91 \\
\hline
 $|1^1s_0, 1^1s_0, 1^1S_0\rangle$  &10.44  &(10.55,\,10.55)   &10.66 \\
 $|1^1p_1, 1^1s_0, 1^3S_1\rangle$  &10.82  &(10.93,\,10.94)   &11.05 \\
 $|1^1d_2, 1^1s_0, 1^5S_2\rangle$  &11.04  &(11.15,\,11.16)   &11.26 \\
 $|1^1f_3, 1^1s_0, 1^7S_3\rangle$  &11.22  &(11.32,\,11.33)   &11.43 \\
 $|1^1g_4, 1^1s_0, 1^9S_4\rangle$  &11.37  &(11.47,\,11.47)   &11.58 \\
 $|1^1h_5, 1^1s_0, 1^{11}S_5\rangle$&11.50 &(11.60,\,11.59)   &11.70 \\
\hline
 $|1^3s_1, 1^1s_0, 1^3S_1\rangle$  &10.47  &(10.58,\,10.59)   &10.69 \\
 $|1^3p_2, 1^1s_0, 1^5S_2\rangle$  &10.83  &(10.94,\,10.95)   &10.06 \\
 $|1^3d_3, 1^1s_0, 1^7S_3\rangle$  &11.05  &(11.16,\,11.17)   &11.27 \\
 $|1^3f_4, 1^1s_0, 1^9S_4\rangle$  &11.22  &(11.33,\,11.33)   &11.44 \\
 $|1^3g_5, 1^1s_0, 1^{11}S_5\rangle$&11.37 &(11.48,\,11.47)   &11.58 \\
 $|1^3h_6, 1^1s_0, 1^{13}S_6\rangle$&11.50 &(11.61,\,11.60)   &11.71 \\
\hline\hline
\end{tabular*}
\end{table*}

\begin{figure*}[!phtb]
\centering
\subfigure[]{\label{subfigure:cfa}\includegraphics[scale=0.45]{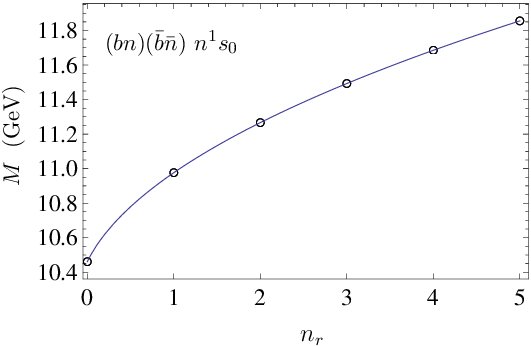}}
\subfigure[]{\label{subfigure:cfa}\includegraphics[scale=0.45]{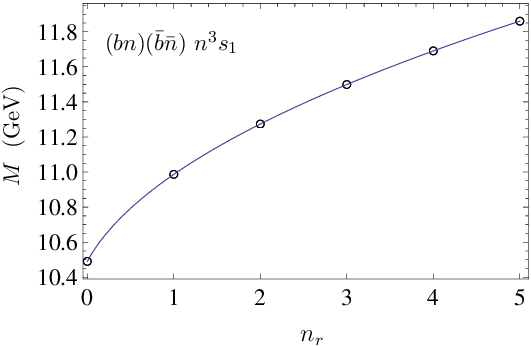}}
\subfigure[]{\label{subfigure:cfa}\includegraphics[scale=0.45]{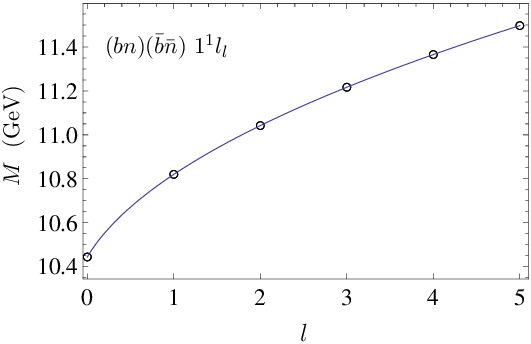}}
\subfigure[]{\label{subfigure:cfa}\includegraphics[scale=0.45]{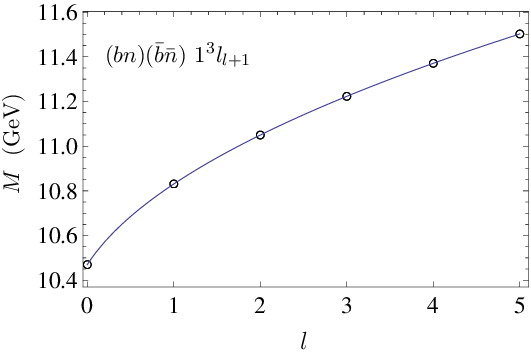}}
\subfigure[]{\label{subfigure:cfa}\includegraphics[scale=0.45]{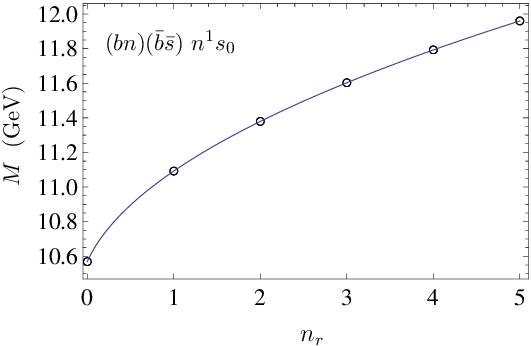}}
\subfigure[]{\label{subfigure:cfa}\includegraphics[scale=0.45]{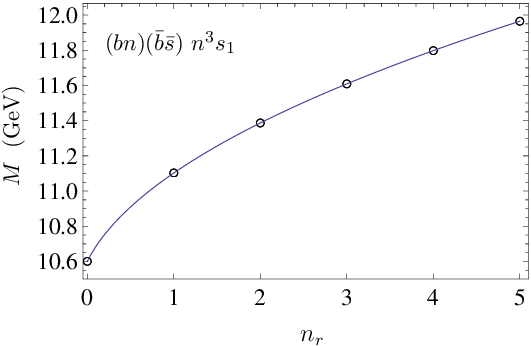}}
\subfigure[]{\label{subfigure:cfa}\includegraphics[scale=0.45]{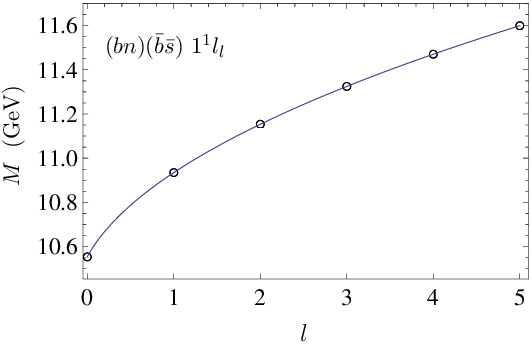}}
\subfigure[]{\label{subfigure:cfa}\includegraphics[scale=0.45]{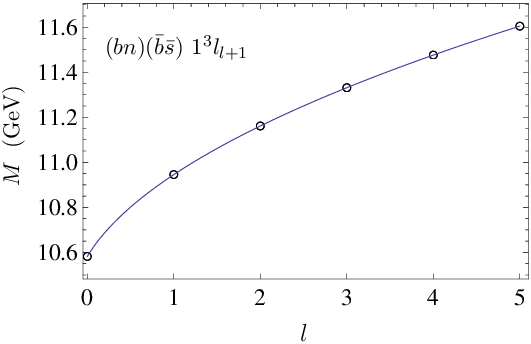}}
\subfigure[]{\label{subfigure:cfa}\includegraphics[scale=0.45]{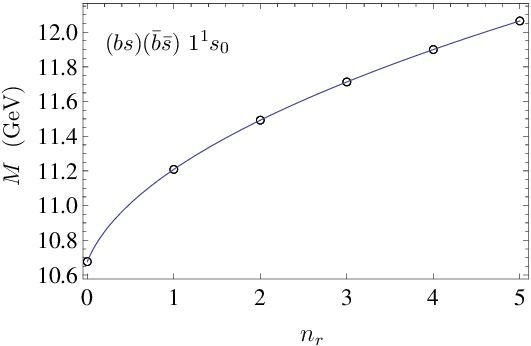}}
\subfigure[]{\label{subfigure:cfa}\includegraphics[scale=0.45]{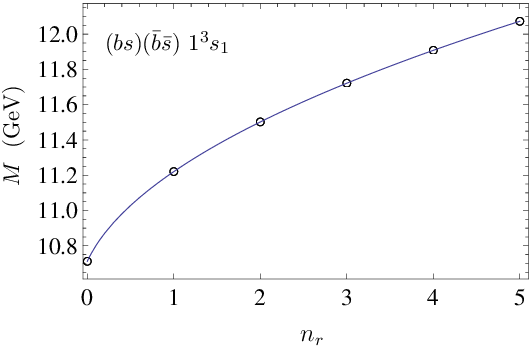}}
\subfigure[]{\label{subfigure:cfa}\includegraphics[scale=0.45]{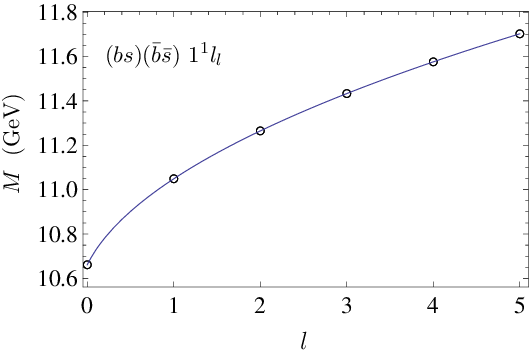}}
\subfigure[]{\label{subfigure:cfa}\includegraphics[scale=0.45]{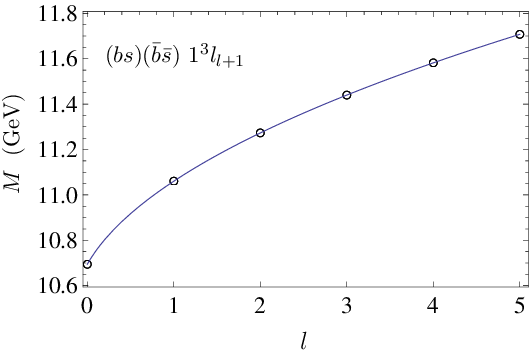}}
\caption{The $\rho$-trajectories for the hidden bottom tetraquarks. Circles indicate the predicted data and the blue lines are the $\rho$-trajectories, see Eq. (\ref{t2qx}). Data are listed in Table \ref{tab:massrho}. $M$ is the bound state mass. $n_r$ and $l$ are the radial and orbital quantum numbers for the $\rho$-mode, respectively. $1^1s_0$ and $1^1l_l$ denote the spin singlet states. $1^3s_1$ and $1^3l_{l+1}$ represent the spin triplet states.}\label{fig:rhort}
\end{figure*}

As discussed in Section \ref{sec:appcfx}, the values of $c_{fx_{\lambda}}$ and $c_{0x_{\lambda}}$ vary with the masses of diquark and antidiquark, $M_{\rho_1}$ and $M_{\rho_2}$, see Eqs. (\ref{t2qx}), (\ref{fitcfxl}) and (\ref{fitcfxnr}). This means that $c_{fx_{\lambda}}$ and $c_{0x_{\lambda}}$ vary with different states of diquark or antidiquark as we discuss the $\rho$-trajectories, even though $\lambda$-mode remains ground state.
Using Eqs. (\ref{t2qx}), (\ref{fitcfxl}) and (\ref{fitcfxnr}) and parameters in Table \ref{tab:parmv}, the masses of the $\rho-$mode excited states are calculated and listed in Table \ref{tab:massrho}.

The mass of the state $|n_1^{2s_1+1}l_{1j_1},n_2^{2s_2+1}l_{2j_2},N^{2j+1}L_J\rangle$ for $(bq)(\bar{b}\bar{q'})$ is equal to that of the state $|n_2^{2s_2+1}l_{2j_2},n_1^{2s_1+1}l_{1j_1},N^{2j+1}L_J\rangle$ for $(bq')(\bar{b}\bar{q})$. For the $\rho-$mode excited states, the mass of the state $|n_1^{2s_1+1}l_{1j_1},n_2^{2s_2+1}l_{2j_2},N^{2j+1}L_J\rangle$ with definite parity and charge parity (for $q=q'$) $1/\sqrt{2}\big[(bq)(\bar{b}\bar{q'})\pm(bq')(\bar{b}\bar{q})\big]$ is equal to that of state
\begin{align}\label{neqtet}
|n_1^{2s_1+1}l_{1j_1},&n_2^{2s_2+1}l_{2j_2},N^{2j+1}L_J\rangle'\nonumber\\
&=\frac{1}{\sqrt{2}}\Big(|n_1^{2s_1+1}l_{1j_1},n_2^{2s_2+1}l_{2j_2},N^{2j+1}L_J\rangle\nonumber\\
&\pm
|n_2^{2s_2+1}l_{2j_2},n_1^{2s_1+1}l_{1j_1},N^{2j+1}L_J\rangle\Big)
\end{align}
for $(bq)(\bar{b}\bar{q'})$ if $n_1^{2s_1+1}l_{1j_1}$ is not identical to $n_2^{2s_2+1}l_{2j_2}$. The corresponding masses should be $M=(M^{\prime}+M^{\prime\prime})/2$, where $M^{\prime}$ and $M^{\prime\prime}$ are listed in Table \ref{tab:massrho}. Therefore, the corresponding tetraquarks will be degenerate in mass. The $\rho$-trajectories are shown in Fig. \ref{fig:rhort}.

The obtained results for the $\lambda$-mode excited states are in agreement with other theoretical predictions, see Table \ref{tab:masscomp}. To our knowledge, the highly radially and orbitally excited states of the $\lambda$-mode states, $|1^1s_0, 1^1s_0, 4^1S_0\rangle$, $|1^1s_0, 1^1s_0, 5^1S_0\rangle$, $|1^1s_0, 1^1s_0, 1^1F_3\rangle$, $|1^1s_0, 1^1s_0, 1^1G_4\rangle$, $|1^1s_0, 1^1s_0, 1^1H_5\rangle$, and the $\rho$-trajectories of the {\hbts} are discussed for the first time.

The observed $T_{b\bar{b}1}(10610)^+$ has been investigated by various theoretical approaches, see Ref. \cite{Zhang:2022hfa} for references. If $T_{b\bar{b}1}(10610)^+$ is a tetraquark, it is the possible candidate for the $|1^3s_1, 1^1s_0, 1^3S_1\rangle'$ state or the $|1^3s_1, 1^3s_1, 1^3S_1\rangle$  state, see Table \ref{tab:expcomp}. The theoretical predictions agree with the experimental data.

\begin{table*}[!phtb]%\color{blue}
\caption{Comparison of theoretical predictions and experimental results (in GeV). $C$ is defined only for $q=q'$. $|n_1^{2s_1+1}l_{1j_1},n_2^{2s_2+1}l_{2j_2},N^{2j+1}L_J\rangle$ is defined in Eq. (\ref{neqtet}).}  \label{tab:expcomp}
\centering
%\rowcolors{1}{blue}{blue}%%
\begin{tabular*}{1.0\textwidth}{@{\extracolsep{\fill}}ccccccc@{}}
\hline\hline
 $J^{PC}$  & Tetraquark & Name &  State candidate  &  Our  & PDG \cite{ParticleDataGroup:2024cfk} & FGS \cite{Faustov:2021hjs}  \\
 &&&$|n_1^{2s_1+1}l_{1j_1},n_2^{2s_2+1}l_{2j_2},N^{2j+1}L_J\rangle$& \\
\hline
 $1^{+-}$  &  $(cn)(\bar{c}\bar{n})$ &  $T_{c\bar{c}1}(3900)^+$ &  $|1^3s_1, 1^1s_0, 1^3S_1\rangle'$ &3.95   &3.8871 &3.871  \\
           &&& $|1^3s_1, 1^3s_1, 1^3S_1\rangle$  &&&3.890\\
          &     & $T_{c\bar{c}1}(4430)^+$ &  $|2^3s_1, 1^1s_0, 1^3S_1\rangle'$ &4.44    &4.478    &   \\
           &&& $|1^3s_1, 2^1s_0, 1^3S_1\rangle'$  &4.51&\\
           &&& $|1^3s_1, 1^1s_0, 2^3S_1\rangle'$  &4.50&&4.431\\
           &&& $|1^3s_1, 1^3s_1, 2^3S_1\rangle$  &&&4.461\\
          & $(cn)(\bar{c}\bar{s})$ & $T_{c\bar{c}\bar{s}1}(4000)^+$ &  $|1^3s_1, 1^1s_0, 1^3S_1\rangle'$ &4.05   &3.988  &3.982  \\
           &&& $|1^3s_1, 1^3s_1, 1^3S_1\rangle$  &&&4.004\\
           &  $(bn)(\bar{b}\bar{n})$ & $T_{b\bar{b}1}(10610)^+$ &  $|1^3s_1, 1^1s_0, 1^3S_1\rangle'$ &10.49   &10.6072 &10.492  \\
           &&& $|1^3s_1, 1^3s_1, 1^3S_1\rangle$  &&&10.494\\
\hline\hline
\end{tabular*}
\end{table*}

\subsection{$\lambda$- and $\rho$-{\trs} for the hidden charm tetraquarks}\label{subsec:rts}

\begin{table*}[!phtb]
\caption{The masses of the $\lambda$-mode excited states of the hidden charm tetraquarks (in ${\gev}$). The notation in Eq. (\ref{tetnot}) is rewritten as $|n_1^{2s_1+1}l_{1j_1},n_2^{2s_2+1}l_{2j_2},N^{2j+1}L_J\rangle$. $n=u,\,d$. Eqs. (\ref{t2qx}), (\ref{fitcfxl}) and (\ref{fitcfxnr}) are used.}  \label{tab:masslambdac}
\centering
\begin{tabular*}{1.0\textwidth}{@{\extracolsep{\fill}}cccc@{}}
\hline\hline
  $|n_1^{2s_1+1}l_{1j_1},n_2^{2s_2+1}l_{2j_2},N^{2j+1}L_J\rangle$        & $(cn)(\bar{c}\bar{n})$  &  $(cn)(\bar{c}\bar{s})$  &  $(cs)(\bar{c}\bar{s})$ \\
\hline
 $|1^1s_0, 1^1s_0, 1^1S_0\rangle$  &3.83    &3.94    &4.05  \\
 $|1^1s_0, 1^1s_0, 2^1S_0\rangle$  &4.39    &4.50    &4.60  \\
 $|1^1s_0, 1^1s_0, 3^1S_0\rangle$  &4.80    &4.90    &5.00  \\
 $|1^1s_0, 1^1s_0, 4^1S_0\rangle$  &5.15    &5.25    &5.34  \\
 $|1^1s_0, 1^1s_0, 5^1S_0\rangle$  &5.46    &5.56    &5.65  \\
\hline
 $|1^1s_0, 1^1s_0, 1^1S_0\rangle$  &3.84    &3.95    &4.05  \\
 $|1^1s_0, 1^1s_0, 1^1P_1\rangle$  &4.23    &4.33    &4.44  \\
 $|1^1s_0, 1^1s_0, 1^1D_2\rangle$  &4.52    &4.63    &4.73  \\
 $|1^1s_0, 1^1s_0, 1^1F_3\rangle$  &4.78    &4.88    &4.98  \\
 $|1^1s_0, 1^1s_0, 1^1G_4\rangle$  &5.01    &5.11    &5.21  \\
 $|1^1s_0, 1^1s_0, 1^1H_5\rangle$  &5.23    &5.32    &5.42  \\
\hline\hline
\end{tabular*}
\end{table*}

\begin{figure*}[!phtb]
\centering
\subfigure[]{\label{subfigure:cfa}\includegraphics[scale=0.6]{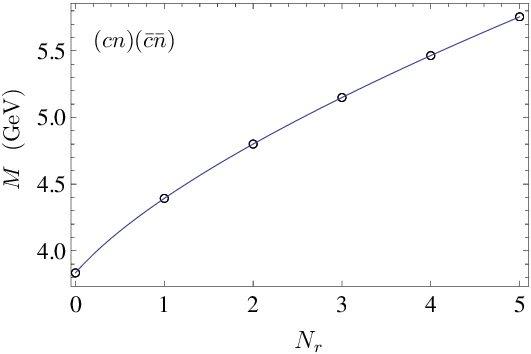}}
\subfigure[]{\label{subfigure:cfa}\includegraphics[scale=0.6]{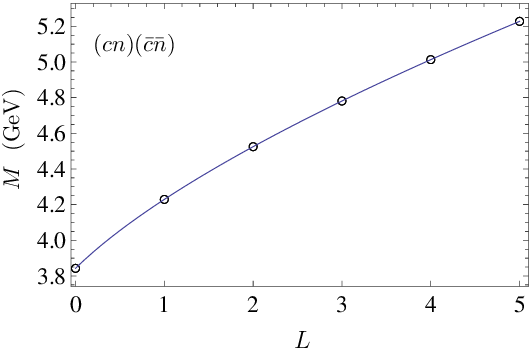}}
\subfigure[]{\label{subfigure:cfa}\includegraphics[scale=0.6]{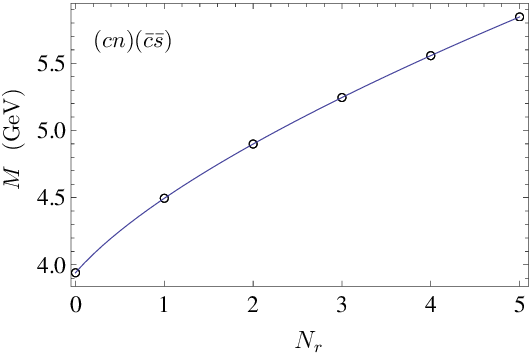}}
\subfigure[]{\label{subfigure:cfa}\includegraphics[scale=0.6]{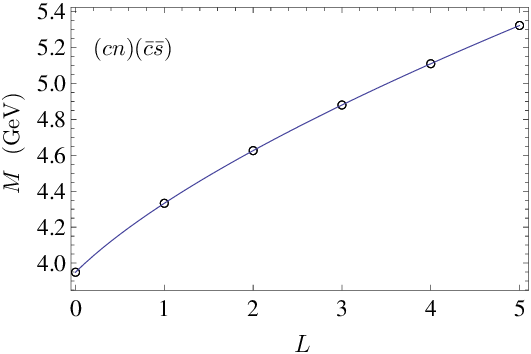}}
\subfigure[]{\label{subfigure:cfa}\includegraphics[scale=0.6]{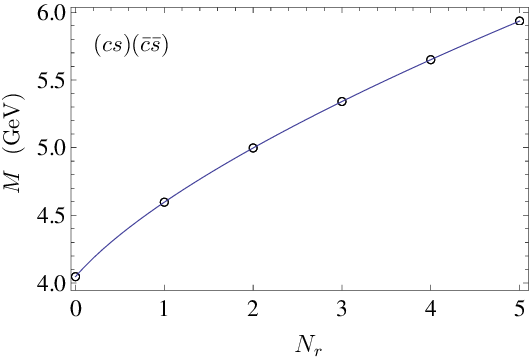}}
\subfigure[]{\label{subfigure:cfa}\includegraphics[scale=0.6]{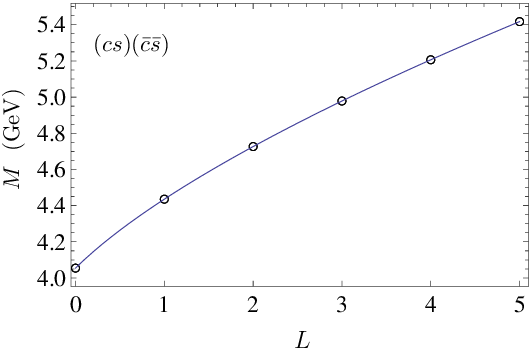}}
\caption{Same as Fig. 2 except for the hidden charm tetraquarks
and Table VI.}\label{fig:tlamc}
\end{figure*}

\begin{table*}[!phtb]
\caption{Comparison of theoretical predictions for the masses of the hidden charm tetraquarks (in GeV). $C$ is defined only for $q=q'$.}  \label{tab:masscompc}
\centering
\begin{tabular*}{1.0\textwidth}{@{\extracolsep{\fill}}ccccccccccc@{}}
\hline\hline
 $J^{PC}$ &  $|n_1^{2s_1+1}l_{1j_1},n_2^{2s_2+1}l_{2j_2},N^{2j+1}L_J\rangle$
          & Tetraquark  &   Our  & FGS \cite{Faustov:2021hjs} & FS \cite{Ferretti:2020ewe} & TR \cite{Tiwari:2022azj} & HK \cite{Hadizadeh:2015cvx} & GLM \cite{Giron:2021sla,Giron:2020fvd} \\
\hline
 $0^{++}$ &  $|1^1s_0, 1^1s_0, 1^1S_0\rangle$ &
                $(cn)(\bar{c}\bar{n})$ &3.83   &3.812   &    &    &  3.739 &3.8419 \\
         &   &  $(cn)(\bar{c}\bar{s})$ &3.94   &3.922   &3.852    &3.955  &  &3.9518  \\
         &   &  $(cs)(\bar{c}\bar{s})$ &4.05   &4.051   &    &4.045    & 3.946 \\
%%%%%%%%%%%%%%
         &  $|1^1s_0, 1^1s_0, 2^1S_0\rangle$
             & $(cn)(\bar{c}\bar{n})$  &4.39   &4.375   &    &    &4.357  \\
         &   &  $(cn)(\bar{c}\bar{s})$ &4.50   &4.481   &4.383    &4.500    &  \\
         &   &  $(cs)(\bar{c}\bar{s})$ &4.60   &4.604   &    &4.620    &4.558  \\
%%%%%%%%%%%%%%%%%%
  $1^{--}$  &$|1^1s_0, 1^1s_0, 1^1P_1\rangle$  &
               $(cn)(\bar{c}\bar{n})$ &4.23   &4.244   &    &    &4.231 &4.2244  \\
        &   &  $(cn)(\bar{c}\bar{s})$ &4.33   &4.350   &4.234    &4.413    &  \\
        &   &  $(cs)(\bar{c}\bar{s})$ &4.44   &4.466   &    &4.556    &4.464  \\
\hline\hline
\end{tabular*}
\end{table*}

The discussions on the hidden charm tetraquarks are the same as those on the hidden bottom tetraquarks. We assume $m_u=m_d$, therefore, the masses for diquarks $(cd)$ and $(cu)$ are equal and it holds for the tetraquarks as well.

When considering the $\lambda$-mode excitations, the scalar diquark and the scalar antidiquark are used. Employing Eqs. (\ref{t2qx}), (\ref{fitcfxl}), (\ref{fitcfxnr}), and parameters in Table \ref{tab:parmv}, the masses of the $\lambda$-mode excited states are estimated, see Table \ref{tab:masslambdac}. The obtained results for the $\lambda$-mode excited states are in agreement with other theoretical predictions, see Table \ref{tab:masscompc}. The $\lambda$-trajectories are shown in Fig. \ref{fig:tlamc}.

When considering the $\rho$-mode excited states, the $\lambda$-mode state is the ground state, $L=0$ and $N_r=0$. The masses of the $\rho$-mode excited states are calculated, see Table \ref{tab:massrhoc}.

\begin{table*}[!htbp]
\caption{The masses of the $\rho$-mode excited states of the hidden charm tetraquarks (in ${\gev}$). The notation in Eq. (\ref{tetnot}) is rewritten as $|n_1^{2s_1+1}l_{1j_1},n_2^{2s_2+1}l_{2j_2},N^{2j+1}L_J\rangle$. $(M^{\prime},\,M^{\prime\prime})$ is for
$\big(|n_1^{2s_1+1}l_{1j_1},n_2^{2s_2+1}l_{2j_2},N^{2j+1}L_J\rangle,$ $|n_2^{2s_2+1}l_{2j_2},n_1^{2s_1+1}l_{1j_1},N^{2j+1}L_J\rangle\big)$. For $(cn)(\bar{c}\bar{n})$ and $(cs)(\bar{c}\bar{s})$, $M'=M^{\prime\prime}$. $n=u,\,d$. Eqations (\ref{t2qx}), (\ref{fitcfxl}), and (\ref{fitcfxnr}) are used.}  \label{tab:massrhoc}
\centering
\begin{tabular*}{1.0\textwidth}{@{\extracolsep{\fill}}cccc@{}}
\hline\hline
  $|n_1^{2s_1+1}l_{1j_1},n_2^{2s_2+1}l_{2j_2},N^{2j+1}L_J\rangle$        & $(cn)(\bar{c}\bar{n})$  &  $(cn)(\bar{c}\bar{s})$  &  $(cs)(\bar{c}\bar{s})$ \\
  &        $M^{\prime}=M^{\prime\prime}$         &  $(M^{\prime},\,M^{\prime\prime})$              & $M^{\prime}=M^{\prime\prime}$   \\
\hline
 $|1^1s_0, 1^1s_0, 1^1S_0\rangle$  &3.83  &(3.94,\,3.94)   &4.05 \\
 $|2^1s_0, 1^1s_0, 1^1S_0\rangle$  &4.40  &(4.50,\,4.56)   &4.67 \\
 $|3^1s_0, 1^1s_0, 1^1S_0\rangle$  &4.70  &(4.80,\,4.87)   &4.97 \\
 $|4^1s_0, 1^1s_0, 1^1S_0\rangle$  &4.93  &(5.03,\,5.11)   &5.21 \\
 $|5^1s_0, 1^1s_0, 1^1S_0\rangle$  &5.12  &(5.23,\,5.31)   &5.41 \\
\hline
 $|1^3s_1, 1^1s_0, 1^3S_1\rangle$  &3.95  &(4.05,\,4.04)   &4.15 \\
 $|2^3s_1, 1^1s_0, 1^3S_1\rangle$  &4.44  &(4.54,\,4.59)   &4.69 \\
 $|3^3s_1, 1^1s_0, 1^3S_1\rangle$  &4.72  &(4.83,\,4.89)   &4.99 \\
 $|4^3s_1, 1^1s_0, 1^3S_1\rangle$  &4.95  &(5.05,\,5.12)   &5.23 \\
 $|5^3s_1, 1^1s_0, 1^3S_1\rangle$  &5.14  &(5.25,\,5.32)   &5.42 \\
\hline
 $|1^1s_0, 1^1s_0, 1^1S_0\rangle$  &3.83  &(3.94,\,3.95)   &4.06 \\
 $|1^1p_1, 1^1s_0, 1^3S_1\rangle$  &4.28  &(4.38,\,4.42)   &4.52 \\
 $|1^1d_2, 1^1s_0, 1^5S_2\rangle$  &4.52  &(4.63,\,4.66)   &4.76 \\
 $|1^1f_3, 1^1s_0, 1^7S_3\rangle$  &4.71  &(4.82,\,4.85)   &4.95 \\
 $|1^1g_4, 1^1s_0, 1^9S_4\rangle$  &4.87  &(4.98,\,5.00)   &5.11 \\
 $|1^1h_5, 1^1s_0, 1^{11}S_5\rangle$&5.01 &(5.12,\,5.14)   &5.25 \\
\hline
 $|1^3s_1, 1^1s_0, 1^3S_1\rangle$  &3.91  &(4.02,\,4.03)   &4.13 \\
 $|1^3p_2, 1^1s_0, 1^5S_2\rangle$  &4.30  &(4.41,\,4.44)   &4.55 \\
 $|1^3d_3, 1^1s_0, 1^7S_3\rangle$  &4.54  &(4.65,\,4.68)   &4.78 \\
 $|1^3f_4, 1^1s_0, 1^9S_4\rangle$  &4.73  &(4.83,\,4.86)   &4.96 \\
 $|1^3g_5, 1^1s_0, 1^{11}S_5\rangle$&4.89 &(4.99,\,5.02)   &5.12 \\
 $|1^3h_6, 1^1s_0, 1^{13}S_6\rangle$&5.03 &(5.13,\,5.16)   &5.26 \\
\hline\hline
\end{tabular*}
\end{table*}

\begin{figure*}[!phtb]
\centering
\subfigure[]{\label{subfigure:cfa}\includegraphics[scale=0.45]{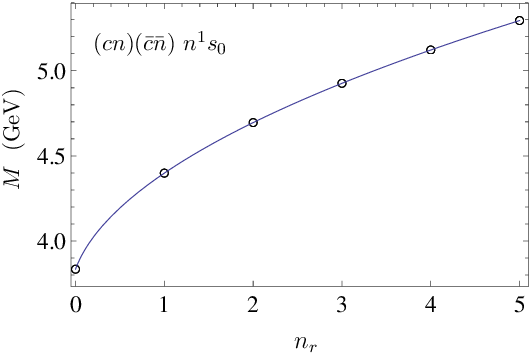}}
\subfigure[]{\label{subfigure:cfa}\includegraphics[scale=0.45]{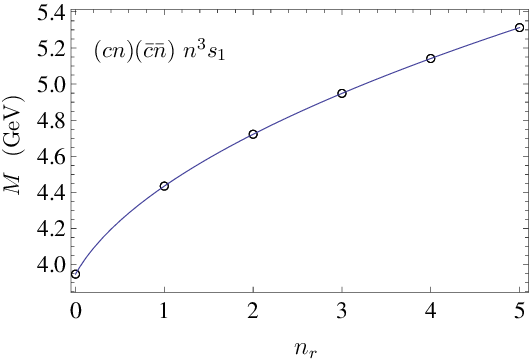}}
\subfigure[]{\label{subfigure:cfa}\includegraphics[scale=0.45]{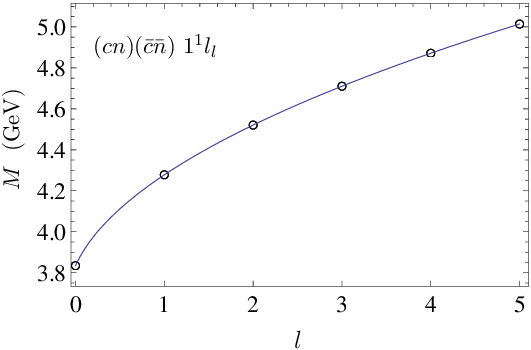}}
\subfigure[]{\label{subfigure:cfa}\includegraphics[scale=0.45]{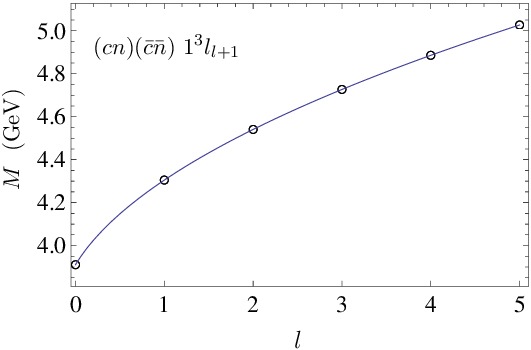}}
\subfigure[]{\label{subfigure:cfa}\includegraphics[scale=0.45]{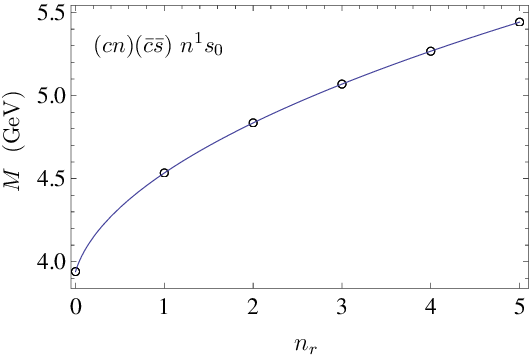}}
\subfigure[]{\label{subfigure:cfa}\includegraphics[scale=0.45]{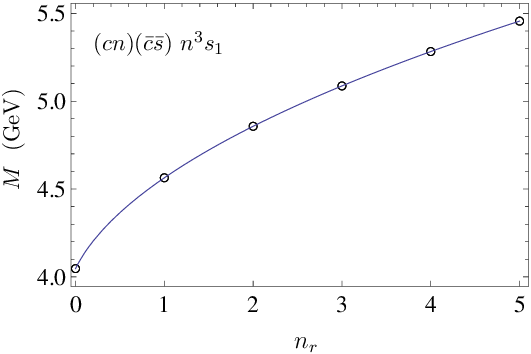}}
\subfigure[]{\label{subfigure:cfa}\includegraphics[scale=0.45]{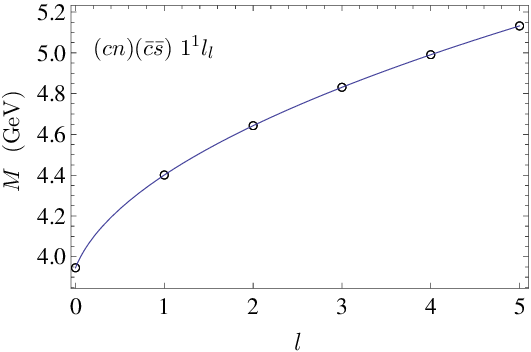}}
\subfigure[]{\label{subfigure:cfa}\includegraphics[scale=0.45]{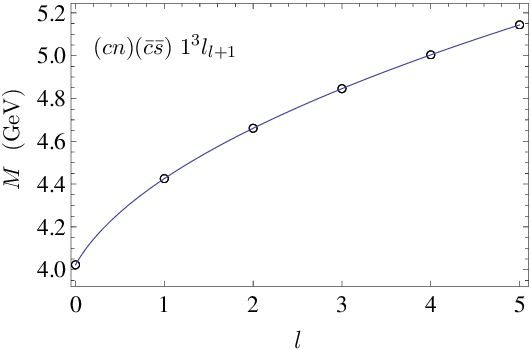}}
\subfigure[]{\label{subfigure:cfa}\includegraphics[scale=0.45]{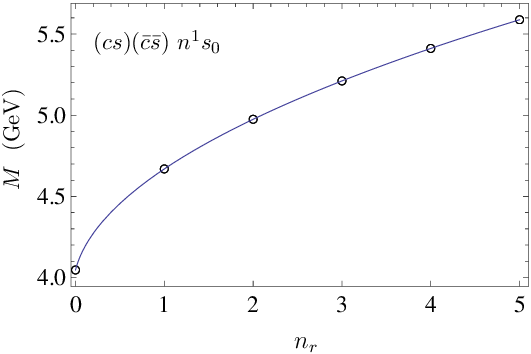}}
\subfigure[]{\label{subfigure:cfa}\includegraphics[scale=0.45]{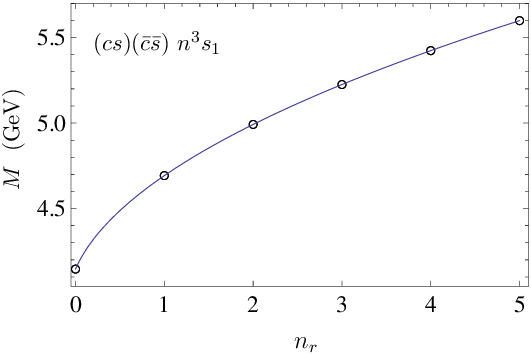}}
\subfigure[]{\label{subfigure:cfa}\includegraphics[scale=0.45]{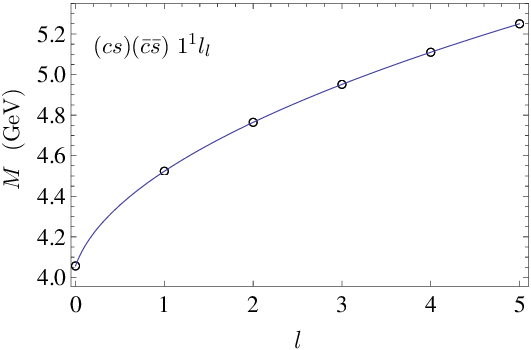}}
\subfigure[]{\label{subfigure:cfa}\includegraphics[scale=0.45]{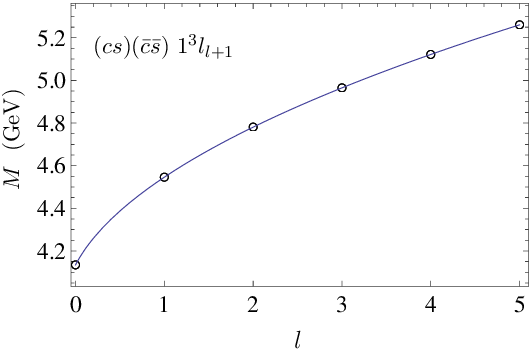}}
\caption{Same as Fig. 3 except for the hidden charm tetraquarks and Table VIII.}\label{fig:rhortc}
\end{figure*}

The mass of the state $|n_1^{2s_1+1}l_{1j_1},n_2^{2s_2+1}l_{2j_2},N^{2j+1}L_J\rangle$ for $(cq)(\bar{c}\bar{q'})$ is equal to that of the state $|n_2^{2s_2+1}l_{2j_2},n_1^{2s_1+1}l_{1j_1},N^{2j+1}L_J\rangle$ for $(cq')(\bar{c}\bar{q})$. For the $\rho-$mode excited states, the mass of the state $|n_1^{2s_1+1}l_{1j_1},n_2^{2s_2+1}l_{2j_2},N^{2j+1}L_J\rangle$ with definite parity and charge parity (for $q=q'$) $1/\sqrt{2}\big[(cq)(\bar{c}\bar{q'})\pm(cq')(\bar{c}\bar{q})\big]$ is equal to that of state
\begin{align}
|n_1^{2s_1+1}l_{1j_1},&n_2^{2s_2+1}l_{2j_2},N^{2j+1}L_J\rangle'\nonumber\\
&=\frac{1}{\sqrt{2}}\Big(|n_1^{2s_1+1}l_{1j_1},n_2^{2s_2+1}l_{2j_2},N^{2j+1}L_J\rangle\nonumber\\
&\pm
|n_2^{2s_2+1}l_{2j_2},n_1^{2s_1+1}l_{1j_1},N^{2j+1}L_J\rangle\Big)\nonumber
\end{align}
for $(cq)(\bar{c}\bar{q'})$ if $n_1^{2s_1+1}l_{1j_1}$ is not identical to $n_2^{2s_2+1}l_{2j_2}$. The corresponding masses should be $M=(M^{\prime}+M^{\prime\prime})/2$, where $M^{\prime}$ and $M^{\prime\prime}$ are listed in Table \ref{tab:massrhoc}. Therefore, the corresponding tetraquarks will be degenerate in mass. The $\rho$-trajectories are shown in Fig. \ref{fig:rhortc}.

In Ref. \cite{Faustov:2021hjs}, $T_{c\bar{c}1}(3900)^+$ and $T_{c\bar{c}\bar{s}1}(4000)^+$ are regarded as $|1^3s_1, 1^1s_0, 1^3S_1\rangle'$ and $|1^3s_1, 1^3s_1, 1^3S_1\rangle'$ states, respectively.
The masses of $T_{c\bar{c}1}(3900)^+$ and $T_{c\bar{c}\bar{s}1}(4000)^+$ coincide with the theoretical predictions for the $1^+$ states, $|1^3s_1, 1^1s_0, 1^3S_1\rangle'$ and $|1^3s_1, 1^3s_1, 1^3S_1\rangle$, see Table \ref{tab:expcomp}.
The observed $T_{c\bar{c}1}(4430)^+$ has the same $J^{PC}$ and quark content as $T_{c\bar{c}1}(3900)^+$. Its mass is $4.478^{+15}_{-18}$ $\gev$ \cite{ParticleDataGroup:2024cfk}.
In Ref. \cite{Hadizadeh:2015cvx}, $T_{c\bar{c}1}(4430)^+$ is regarded as the first radially $\lambda$-excited state $|1^3s_1, 1^3s_1, 2^3S_1\rangle$ of $(cn)(\bar{c}\bar{n})$. In Ref. \cite{Faustov:2021hjs}, $T_{c\bar{c}1}(4430)^+$ is taken as $|1^3s_1, 1^3s_1, 2^3S_1\rangle$ state or $|1^3s_1, 1^1s_0, 2^3S_1\rangle'$ state.
The mass of $T_{c\bar{c}1}(4430)^+$ also approximates the mass of the $|2^3s_1, 1^1s_0, 1^3S_1\rangle'$ state and the $|1^3s_1, 2^1s_0, 1^3S_1\rangle'$ state, see Table \ref{tab:massrhoc}. The states $|2^3s_1, 1^1s_0, 1^3S_1\rangle'$ and $|1^3s_1, 2^1s_0, 1^3S_1\rangle'$ are the $\rho$-excited states. We conclude that $T_{c\bar{c}1}(4430)^+$ is possibly the first radial $\lambda$-excitation or the first radial $\rho$-excitation.

\subsection{Discussions}

By employing the {\rt} formulas in (\ref{t2qx}), the masses for the more highly excited states can be calculated easily. We argue that the calculation of the masses of highly excited state is instructive although the highly excited states will be unstable. This is because various models or approaches to calculate the masses of tetraquarks often use different parameter values. These models, despite using different parameter values, often give agreeable predictions for the ground state and the lower excited states. The theoretical predictions for the highly excited states are expected to show differences, which will be instructive.

The {\rts} take different form and behave differently in various energy regions \cite{Chen:2022flh,Chen:2021kfw}. The hidden charm tetraquark case is similar to the hidden bottom tetraquark case.
Both the diquark $(Qq)$ and antidiquark $(\bar{Q}\bar{q}')$ are the heavy-light systems; therefore, the $\rho$-trajectories of the {\hbts} behave as $M{\sim}x_{\rho}^{1/2}$, see Eq. (\ref{t2qx}). For the $\lambda$-mode, the tetraquark $(Qq)(\bar{Q}\bar{q}')$ is the heavy-heavy system; hence, the $\lambda$-trajectories of the {\hbts} behave as $M{\sim}x_{\lambda}^{2/3}$.

According to Eq. (\ref{t2qx}), there are three series of {\rts}: two series of $\rho$-trajectories and one series of $\lambda$-trajectories. Correspondingly, there are three series of masses. When $q=q'$, there is degeneracy in the $\rho$-trajectories, corresponding to the degeneracy in masses of the $\rho$-excited states.

For the {\hbts}, both the $\lambda$-trajectories and the $\rho$-trajectories are concave downward in the $(M^2,\,x)$ planes if the confining potential is linear. In the subsection \ref{subsec:rtdisc}, we show that for all tetraquarks, not only the $\lambda$-trajectories but also the $\rho$-trajectories are concave downward in the $(M^2,\,x)$ planes when the masses of the light constituents are considered and the confining potential is linear.

\section{Conclusions}\label{sec:conc}

In this work, we obtain the {\rt} formulas for the {\hbts} {\twhf} by using the {\rt} formula for the heavy-heavy systems and the newly proposed {\rt} relation for the heavy-light diquarks. We apply the obtained {\rt} formulas to investigate the {\twh}.
The masses of the $\lambda$-mode excited and $\rho$-mode excited tetraquarks are estimated. The calculated results are in agreement with other theoretical predictions. Both the $\lambda$-trajectories and $\rho$-trajectories are discussed.

For the {\hbts}, there are three series of the {\rts}: two series of $\rho$-trajectories and one series of $\lambda$-trajectories. They exhibit different behaviors.
The $\rho$-trajectories behave as $M{\sim}x_{\rho}^{1/2}$ because the $\rho$-mode excitations are in the heavy-light diquark $(Qq)$ and antidiquark $(\bar{Q}\bar{q}')$. In contrast, the $\lambda$-trajectories behave as $M{\sim}x_{\lambda}^{2/3}$ because the $\lambda$-mode excitations occur between the diquark and the antidiquark, with the diquark $(Qq)$ and antidiquark $(\bar{Q}\bar{q}')$ forming a heavy-heavy system $(Qq)(\bar{Q}\bar{q}')$.

Moreover, the behaviors of the {\rts} for various tetraquarks are investigated based on the spinless Salpeter equation. It is shown that both the $\lambda$-trajectories and the $\rho$-trajectories for all tetraquarks are concave downward in the $(M^2,\,x)$ plane when the masses of the light constituents are considered and the confining potential is linear.

\section*{Acknowledgments}
We are very grateful to the anonymous referees for the valuable comments and suggestions.

\appendix
\section{States of diquarks}

\begin{table}[!phtb]
\caption{The completely antisymmetric states for the diquarks in {\cltb} \cite{Feng:2023txx}. $j$ is the spin of the diquark {\dqs}, $s$ denotes the total spin of two quarks, $l$ represents the orbital angular momentum. $n=n_r+1$, $n_r$ is the radial quantum number, $n_r=0,1,2,\cdots$. }  \label{tab:dqstates}
\centering
\begin{tabular*}{0.49\textwidth}{@{\extracolsep{\fill}}ccccc@{}}
\hline\hline
 Spin of diquark & Parity  &  Wave state  &  Configuration    \\
( $j$ )          & $(j^P)$ & $(n^{2s+1}l_j)$  \\
\hline
j=0              & $0^+$   & $n^1s_0$         & $[qq']^{{\cltba}}_{n^1s_0}$ \\
                 & $0^-$   & $n^3p_0$         & $[qq']^{{\cltba}}_{n^3p_0}$       \\
j=1              & $1^+$   & $n^3s_1$, $n^3d_1$   & $\{qq'\}^{{\cltba}}_{n^3s_1}$,\;    $\{qq'\}^{{\cltba}}_{n^3d_1}$\\
                 & $1^-$   & $n^1p_1$, $n^3p_1$   &
$\{qq'\}^{{\cltba}}_{n^1p_1}$,\; $[qq']^{{\cltba}}_{n^3p_1}$ \\
j=2              & $2^+$   & $n^1d_2$, $n^3d_2$         &  $[qq']^{{\cltba}}_{n^1d_2}$,\; $\{qq'\}^{{\cltba}}_{n^3d_2}$\\
                 & $2^-$   & $n^3p_2$, $n^3f_2$       &
 $[qq']^{{\cltba}}_{n^3p_2}$,\; $[qq']^{{\cltba}}_{n^3f_2}$       \\
$\cdots$         & $\cdots$ & $\cdots$               & $\cdots$  \\
\hline\hline
\end{tabular*}
\end{table}

In this section, we list the completely antisymmetric states for the diquarks in {\cltb} \cite{Feng:2023txx}, see Table IX.

\section{States of tetraquarks}
The states of tetraquark in the diquark picture are listed in Table \ref{tab:tqstates}.

\begin{table*}[!phtb]
\caption{The states of tetraquark in the diquark picture. The diquark notation is in \ref{tab:dqstates}. Here, $q$ and $q'$ represent both the light quarks and the heavy quarks. The superscript $1_c$ is omitted.}  \label{tab:tqstates}
\centering
\begin{tabular*}{\textwidth}{@{\extracolsep{\fill}}ccccc@{}}
\hline\hline
$J^P$ & $(l_1,l_2,L)$  &  Configuration \\
\hline
$0^+$ & $(0,0,0)$ &
$\left([qq']^{{\cltba}}_{n_1^1s_0}
[\bar{q}\bar{q}']^{3_c}_{n_2^1s_0}\right)_{N^1S_0}$,\;
$\left(\{qq'\}^{{\cltba}}_{n_1^3s_1}
\{\bar{q}\bar{q}'\}^{3_c}_{n_2^3s_1}\right)_{N^1S_0}$,\\
%%%%%%%%%%%
& $(1,1,0)$ &
$\left([qq']^{{\cltba}}_{n_1^3p_0}
[\bar{q}\bar{q}']^{3_c}_{n_2^3p_0}\right)_{N^1S_0}$,\;
$\left(\{qq'\}^{{\cltba}}_{n_1^1p_1}
\{\bar{q}\bar{q}'\}^{3_c}_{n_2^1p_1}\right)_{N^1S_0}$,\;
$\left([qq']^{{\cltba}}_{n_1^3p_1}
[\bar{q}\bar{q}']^{3_c}_{n_2^3p_1}\right)_{N^1S_0}$,\;
$\left([qq']^{{\cltba}}_{n_1^3p_2}
[\bar{q}\bar{q}']^{3_c}_{n_2^3p_2}\right)_{N^1S_0}$,\\
& &
$\left(\{qq'\}^{{\cltba}}_{n_1^1p_1}
[\bar{q}\bar{q}']^{3_c}_{n_2^3p_1}\right)_{N^1S_0}$,\;
$\left([qq']^{{\cltba}}_{n_1^3p_1}
\{\bar{q}\bar{q}'\}^{3_c}_{n_2^1p_1}\right)_{N^1S_0}$\\
%%%%%%%%%%%
& $(0,1,1)$ &
$\left([qq']^{{\cltba}}_{n_1^1s_0}
\{\bar{q}\bar{q}'\}^{3_c}_{n_2^1p_1}\right)_{N^3P_0}$,\;
$\left([qq']^{{\cltba}}_{n_1^1s_0}
[\bar{q}\bar{q}']^{3_c}_{n_2^3p_1}\right)_{N^3P_0}$,\;
$\left(\{qq'\}^{{\cltba}}_{n_1^3s_1}
[\bar{q}\bar{q}']^{3_c}_{n_2^3p_0}\right)_{N^3P_0}$,\\
& &
$\left(\{qq'\}^{{\cltba}}_{n_1^3s_1}
\{\bar{q}\bar{q}'\}^{3_c}_{n_2^1p_1}\right)_{N^3P_0}$,\;
$\left(\{qq'\}^{{\cltba}}_{n_1^3s_1}
[\bar{q}\bar{q}']^{3_c}_{n_2^3p_1}\right)_{N^3P_0}$,\;
$\left(\{qq'\}^{{\cltba}}_{n_1^3s_1}
[\bar{q}\bar{q}']^{3_c}_{n_2^3p_2}\right)_{N^3P_0}$\\
%%%%%%%%%%%
& $(1,0,1)$ &
$\left(\{qq'\}^{3_c}_{n_1^1p_1}
[\bar{q}\bar{q}']^{{\cltba}}_{n_2^1s_0}\right)_{N^3P_0}$,\;
$\left([qq']^{3_c}_{n_1^3p_1}
[\bar{q}\bar{q}']^{{\cltba}}_{n_2^1s_0}\right)_{N^3P_0}$,\;
$\left([qq']^{3_c}_{n_1^3p_0}
\{\bar{q}\bar{q}'\}^{{\cltba}}_{n_2^3s_1}\right)_{N^3P_0}$,\\
& &
$\left(\{qq'\}^{3_c}_{n_1^1p_1}
\{\bar{q}\bar{q}'\}^{{\cltba}}_{n_2^3s_1}\right)_{N^3P_0}$,\;
$\left([qq']^{3_c}_{n_1^3p_1}
\{\bar{q}\bar{q}'\}^{{\cltba}}_{n_2^3s_1}\right)_{N^3P_0}$,\;
$\left([qq']^{3_c}_{n_1^3p_2}
\{\bar{q}\bar{q}'\}^{{\cltba}}_{n_2^3s_1}\right)_{N^3P_0}$\\
&$\cdots$ &$\cdots$ \\
%%%%%%%%%%%%%%%%%%%%%%%%%%%%%%%%%%%%%%%%%%%%%%%%%%%%%%%%%%%%%%%%
$0^-$ & $(0,0,1)$ &
$\left([qq']^{{\cltba}}_{n_1^1s_0}
\{\bar{q}\bar{q}'\}^{3_c}_{n_2^3s_1}\right)_{N^3P_0}$,\;
$\left(\{qq'\}^{3_c}_{n_1^3s_1}
[\bar{q}\bar{q}']^{{\cltba}}_{n_2^1s_0}\right)_{N^3P_0}$,\;
$\left(\{qq'\}^{{\cltba}}_{n_1^3s_1}
\{\bar{q}\bar{q}'\}^{3_c}_{n_2^3s_1}\right)_{N^3P_0}$,\\
%%%%%%%%%%%
& $(1,0,0)$ &
$\left([qq']^{{\cltba}}_{n_1^3p_0}
[\bar{q}\bar{q}']^{3_c}_{n_2^1s_0}\right)_{N^1S_0}$,\;
$\left(\{qq'\}^{{\cltba}}_{n_1^1p_1}
\{\bar{q}\bar{q}'\}^{3_c}_{n_2^3s_1}\right)_{N^1S_0}$,\;
$\left([qq']^{{\cltba}}_{n_1^3p_1}
\{\bar{q}\bar{q}'\}^{3_c}_{n_2^3s_1}\right)_{N^1S_0}$,\\
%%%%%%%%%%%
& $(0,1,0)$ &
$\left([qq']^{3_c}_{n_1^1s_0}
[\bar{q}\bar{q}']^{{\cltba}}_{n_2^3p_0}\right)_{N^1S_0}$,\;
$\left(\{qq'\}^{3_c}_{n_1^3s_1}
\{\bar{q}\bar{q}'\}^{{\cltba}}_{n_2^1p_1}\right)_{N^1S_0}$,\;
$\left(\{qq'\}^{3_c}_{n_1^3s_1}
[\bar{q}\bar{q}']^{{\cltba}}_{n_2^3p_1}\right)_{N^1S_0}$,\\
&$\cdots$ &$\cdots$ \\
%%%%%%%%%%%%%%%%%%%%%%%%%%%%%%%%%%%%%%%%%%%%%%%%%%%%%%%%%%%%%
$1^+$ & $(0,0,0)$ &
$\left([qq']^{{\cltba}}_{n_1^1s_0}
\{\bar{q}\bar{q}'\}^{3_c}_{n_2^3s_1}\right)_{N^3S_1}$,\;
$\left(\{qq'\}^{{\cltba}}_{n_1^3s_1}
[\bar{q}\bar{q}']^{3_c}_{n_2^1s_0}\right)_{N^3S_1}$,\;
$\left(\{qq'\}^{{\cltba}}_{n_1^3s_1}
\{\bar{q}\bar{q}'\}^{3_c}_{n_2^3s_1}\right)_{N^3S_1}$,\\
%%%%%%%%%%%
& $(1,1,0)$ &
$\left([qq']^{{\cltba}}_{n_1^3p_0}
\{\bar{q}\bar{q}'\}^{3_c}_{n_2^1p_1}\right)_{N^3S_1}$,\;
$\left([qq']^{{\cltba}}_{n_1^3p_0}
[\bar{q}\bar{q}']^{3_c}_{n_2^3p_1}\right)_{N^3S_1}$,\\
&    &
$\left(\{qq'\}^{{\cltba}}_{n_1^1p_1}
[\bar{q}\bar{q}']^{3_c}_{n_2^3p_0}\right)_{N^3S_1}$,\;
$\left(\{qq'\}^{{\cltba}}_{n_1^1p_1}
\{\bar{q}\bar{q}'\}^{3_c}_{n_2^1p_1}\right)_{N^3S_1}$,\;
$\left(\{qq'\}^{{\cltba}}_{n_1^1p_1}
[\bar{q}\bar{q}']^{3_c}_{n_2^3p_1}\right)_{N^3S_1}$,\;
$\left(\{qq'\}^{{\cltba}}_{n_1^1p_1}
[\bar{q}\bar{q}']^{3_c}_{n_2^3p_2}\right)_{N^3S_1}$,\\
& &
$\left([qq']^{{\cltba}}_{n_1^3p_1}
[\bar{q}\bar{q}']^{3_c}_{n_2^3p_0}\right)_{N^3S_1}$,\;
$\left([qq']^{{\cltba}}_{n_1^3p_1}
\{\bar{q}\bar{q}'\}^{3_c}_{n_2^1p_1}\right)_{N^3S_1}$,\;
$\left([qq']^{{\cltba}}_{n_1^3p_1}
[\bar{q}\bar{q}']^{3_c}_{n_2^3p_1}\right)_{N^3S_1}$,\;
$\left([qq']^{{\cltba}}_{n_1^3p_1}
[\bar{q}\bar{q}']^{3_c}_{n_2^3p_2}\right)_{N^3S_1}$,\\
&&
$\left([qq']^{{\cltba}}_{n_1^3p_2}
\{\bar{q}\bar{q}'\}^{3_c}_{n_2^1p_1}\right)_{N^3S_1}$,\;
$\left([qq']^{{\cltba}}_{n_1^3p_2}
[\bar{q}\bar{q}']^{3_c}_{n_2^3p_1}\right)_{N^3S_1}$,\;
$\left([qq']^{{\cltba}}_{n_1^3p_2}
[\bar{q}\bar{q}']^{3_c}_{n_2^3p_2}\right)_{N^3S_1}$,\\
%%%%%%%%%%%
& $(0,1,1)$ &
$\left([qq']^{{\cltba}}_{n_1^1s_0}
\{\bar{q}\bar{q}'\}^{3_c}_{n_2^1p_1}\right)_{N^3P_1}$,\;
$\left([qq']^{{\cltba}}_{n_1^1s_0}
[\bar{q}\bar{q}']^{3_c}_{n_2^3p_1}\right)_{N^3P_1}$,\;
$\left(\{qq'\}^{{\cltba}}_{n_1^3s_1}
[\bar{q}\bar{q}']^{3_c}_{n_2^3p_0}\right)_{N^3P_1}$,\\
& &
$\left(\{qq'\}^{{\cltba}}_{n_1^3s_1}
\{\bar{q}\bar{q}'\}^{3_c}_{n_2^1p_1}\right)_{N^3P_1}$,\;
$\left(\{qq'\}^{{\cltba}}_{n_1^3s_1}
[\bar{q}\bar{q}']^{3_c}_{n_2^3p_1}\right)_{N^3P_1}$,\;
$\left(\{qq'\}^{{\cltba}}_{n_1^3s_1}
[\bar{q}\bar{q}']^{3_c}_{n_2^3p_2}\right)_{N^3P_1}$\\
& &
$\left(\{qq'\}^{{\cltba}}_{n_1^3s_1}
\{\bar{q}\bar{q}'\}^{3_c}_{n_2^1p_1}\right)_{N^5P_1}$,\;
$\left(\{qq'\}^{{\cltba}}_{n_1^3s_1}
[\bar{q}\bar{q}']^{3_c}_{n_2^3p_1}\right)_{N^5P_1}$,\;
$\left(\{qq'\}^{{\cltba}}_{n_1^3s_1}
[\bar{q}\bar{q}']^{3_c}_{n_2^3p_2}\right)_{N^5P_1}$\\
%%%%%%%%%%%
& $(1,0,1)$ &
$\left(\{qq'\}^{3_c}_{n_1^1p_1}
[\bar{q}\bar{q}']^{{\cltba}}_{n_2^1s_0}\right)_{N^3P_1}$,\;
$\left([qq']^{3_c}_{n_1^3p_1}
[\bar{q}\bar{q}']^{{\cltba}}_{n_2^1s_0}\right)_{N^3P_1}$,\;
$\left([qq']^{3_c}_{n_1^3p_0}
\{\bar{q}\bar{q}'\}^{{\cltba}}_{n_2^3s_1}\right)_{N^3P_1}$,\\
& &
$\left(\{qq'\}^{3_c}_{n_1^1p_1}
\{\bar{q}\bar{q}'\}^{{\cltba}}_{n_2^3s_1}\right)_{N^3P_1}$,\;
$\left([qq']^{3_c}_{n_1^3p_1}
\{\bar{q}\bar{q}'\}^{{\cltba}}_{n_2^3s_1}\right)_{N^3P_1}$,\;
$\left([qq']^{3_c}_{n_1^3p_2}
\{\bar{q}\bar{q}'\}^{{\cltba}}_{n_2^3s_1}\right)_{N^3P_1}$\\
& &
$\left(\{qq'\}^{3_c}_{n_1^1p_1}
\{\bar{q}\bar{q}'\}^{{\cltba}}_{n_2^3s_1}\right)_{N^5P_1}$,\;
$\left([qq']^{3_c}_{n_1^3p_1}
\{\bar{q}\bar{q}'\}^{{\cltba}}_{n_2^3s_1}\right)_{N^5P_1}$,\;
$\left([qq']^{3_c}_{n_1^3p_2}
\{\bar{q}\bar{q}'\}^{{\cltba}}_{n_2^3s_1}\right)_{N^5P_1}$\\
&$\cdots$ &$\cdots$ \\
%%%%%%%%%%%%%%%%%%%%%%%%%%%%%%%%%%%%%%%%%%%%%%%%%%%%%%%%%%%%%
$1^-$ & $(0,0,1)$ &
$\left([qq']^{{\cltba}}_{n_1^1s_0}
\{\bar{q}\bar{q}'\}^{3_c}_{n_2^3s_1}\right)_{N^3P_1}$,\;
$\left(\{qq'\}^{3_c}_{n_1^3s_1}
[\bar{q}\bar{q}']^{{\cltba}}_{n_2^1s_0}\right)_{N^3P_1}$,\;
$\left(\{qq'\}^{{\cltba}}_{n_1^3s_1}
\{\bar{q}\bar{q}'\}^{3_c}_{n_2^3s_1}\right)_{N^3P_1}$,\\
&&
$\left([qq']^{3_c}_{n_1^1s_0}
[\bar{q}\bar{q}']^{{\cltba}}_{n_2^1s_0}\right)_{N^1P_1}$,\;
$\left(\{qq'\}^{{\cltba}}_{n_1^3s_1}
\{\bar{q}\bar{q}'\}^{3_c}_{n_2^3s_1}\right)_{N^1P_1}$,\;
$\left(\{qq'\}^{{\cltba}}_{n_1^3s_1}
\{\bar{q}\bar{q}'\}^{3_c}_{n_2^3s_1}\right)_{N^5P_1}$,\\
%%%%%%%%%%%
& $(1,0,0)$ &
$\left([qq']^{{\cltba}}_{n_1^3p_0}
\{\bar{q}\bar{q}'\}^{3_c}_{n_2^3s_1}\right)_{N^3S_1}$,\;
$\left(\{qq'\}^{{\cltba}}_{n_1^1p_1}
\{\bar{q}\bar{q}'\}^{3_c}_{n_2^3s_1}\right)_{N^3S_1}$,\;
$\left(\{qq'\}^{{\cltba}}_{n_1^1p_1}
[\bar{q}\bar{q}']^{3_c}_{n_2^1s_0}\right)_{N^3S_1}$,\\
&&
$\left([qq']^{{\cltba}}_{n_1^3p_1}
\{\bar{q}\bar{q}'\}^{3_c}_{n_2^3s_1}\right)_{N^3S_1}$,\;
$\left([qq']^{{\cltba}}_{n_1^3p_1}
[\bar{q}\bar{q}']^{3_c}_{n_2^1s_0}\right)_{N^3S_1}$,\;
$\left([qq']^{{\cltba}}_{n_1^3p_2}
\{\bar{q}\bar{q}'\}^{3_c}_{n_2^3s_1}\right)_{N^3S_1}$\\
%%%%%%%%%%%
& $(0,1,0)$ &
$\left(\{qq'\}^{3_c}_{n_1^3s_1}
[\bar{q}\bar{q}']^{{\cltba}}_{n_2^3p_0}\right)_{N^3S_1}$,\;
$\left(\{qq'\}^{3_c}_{n_1^3s_1}
\{\bar{q}\bar{q}'\}^{{\cltba}}_{n_2^1p_1}\right)_{N^3S_1}$,\;
$\left([qq']^{3_c}_{n_1^1s_0}
\{\bar{q}\bar{q}'\}^{{\cltba}}_{n_2^1p_1}\right)_{N^3S_1}$,\\
&&
$\left(\{qq'\}^{3_c}_{n_1^3s_1}
[\bar{q}\bar{q}']^{{\cltba}}_{n_2^3p_1}\right)_{N^3S_1}$,\;
$\left([qq']^{3_c}_{n_1^1s_0}
[\bar{q}\bar{q}']^{{\cltba}}_{n_2^3p_1}\right)_{N^3S_1}$,\;
$\left(\{qq'\}^{3_c}_{n_1^3s_1}
[\bar{q}\bar{q}']^{{\cltba}}_{n_2^3p_2}\right)_{N^3S_1}$\\
&$\cdots$ &$\cdots$ \\
\hline\hline
\end{tabular*}
\end{table*}

\section{{\rt} behaviors for various tetraquarks}\label{app:c}
In this section, the {\rt} behaviors for various tetraquarks are investigated.

\subsection{Spinless Salpeter equation}
The spinless Salpeter equation \cite{Godfrey:1985xj,Ferretti:2019zyh,Bedolla:2019zwg,Durand:1981my,Durand:1983bg,Lichtenberg:1982jp,Jacobs:1986gv} reads as
\begin{eqnarray}\label{qsse}
M\Psi_{d,t}({\bfr})=\left(\omega_1+\omega_2\right)\Psi_{d,t}({\bfr})+V_{d,t}\Psi_{d,t}({\bfr}),
\end{eqnarray}
where $M$ is the bound state mass (diquark or tetraquark). $\Psi_{d,t}({\bfr})$ are the diquark wave function and the tetraquark wave function, respectively. $V_{d,t}$ denotes the diquark potential and the tetraquark potential, respectively (see Eq. (\ref{potv})). $\omega_1$ is the relativistic energy of constituent $1$ (quark or diquark), and $\omega_2$ is of constituent $2$ (quark or antidiquark),
\bea\label{omega}
\omega_i=\sqrt{m_i^2+{\bf p}^2}=\sqrt{m_i^2-\Delta}\;\; (i=1,2).
\eea
$m_1$ and $m_2$ are the effective masses of constituent $1$ and $2$, respectively.

When using diquark in multiquark systems, the interactions between quark and quark, diquark and quark, diquark and diquark are needed. In Refs. \cite{Faustov:2021hjs}, these interactions are constructed with the help of the off-mass-shell scattering amplitude, which is projected onto the positive energy states.
The interactions also can be established by expanding the interactions of the quark-antiquark system to the quark-quark system, and then to the diquark-antidiquark systems or the diquark-quark systems \cite{Lundhammar:2020xvw}.
Furthermore, the effect of the finite size of diquark is treated differently. In Refs. \cite{Faustov:2021hjs}, the size of diquark is taken into account through corresponding form factors. At times, diquark is taken as being pointlike \cite{Ferretti:2019zyh,Lundhammar:2020xvw}, and we use this approximation in the present work.

Following Refs. \cite{Ferretti:2019zyh,Bedolla:2019zwg,Ferretti:2011zz,Eichten:1974af}, we employ the potential
\begin{align}\label{potv}
V_{d,t}&=-\frac{3}{4}\left[V_c+{\sigma}r+C\right]
\left({\bf{F}_i}\cdot{\bf{F}_j}\right)_{d,t},
\end{align}
where $V_c\propto{1/r}$ is a color Coulomb potential or a Coulomb-like potential due to one-gluon-exchange. $\sigma$ is the string tension. $C$ is a fundamental parameter \cite{Gromes:1981cb,Lucha:1991vn}. The part in the bracket is the Cornell potential \cite{Eichten:1974af}. ${\bf{F}_i}\cdot{\bf{F}_j}$ is the color-Casimir,
\bea\label{mrcc}
\langle{(\bf{F}_i}\cdot{\bf{F}_j})_{d}\rangle=-\frac{2}{3},\quad
\langle{(\bf{F}_i}\cdot{\bf{F}_j})_{t}\rangle=-\frac{4}{3}.
\eea

\subsection{{\rt} relations for various systems}
For the heavy-heavy systems, $m_{1},m_2{\gg}{\bfpa}$, Eq. (\ref{qsse}) reduces to
\begin{eqnarray}\label{qssenrr}
M\Psi_{d,t}({\bfr})&=&\left[(m_1+m_2)+\frac{{\bfp}^2}{2\mu}\right]\Psi_{d,t}({\bfr})\nonumber\\
&&+V_{d,t}\Psi_{d,t}({\bfr}),
\end{eqnarray}
where
\bea\label{rdmu}
\mu=m_1m_2/(m_1+m_2).
\eea
By employing the Bohr-Sommerfeld quantization approach \cite{Brau:2000st} and using Eqs. (\ref{potv}) and (\ref{qssenrr}), we obtain the parametrized relation  \cite{Chen:2022flh,Chen:2021kfw}
\bea\label{massform}
M=m_R+\beta_x(x+c_{0x})^{2/3}\,\,(x=l,\,n_r,\,L,\,N_r),
\eea
with
\bea\label{parabm}
\beta_x=c_{fx}c_xc_c,\quad m_R=m_1+m_2+C',
\eea
where
\bea\label{cprime}
C'=\left\{\begin{array}{cc}
C/2, & \text{diquarks}, \\
C, & \text{tetraquarks}.
\end{array}\right.
\eea
\bea\label{sigma}
\sigma'=\left\{\begin{array}{cc}
\sigma/2, & \text{diquarks}, \\
\sigma, & \text{tetraquarks}.
\end{array}\right.
\eea
$c_{x}$ and $c_c$ are
\bea\label{cxcons}
c_c=\left(\frac{\sigma'^2}{\mu}\right)^{1/3},\quad c_{l,L}=\frac{3}{2},\quad c_{n_r,N_r}=\frac{\left(3\pi\right)^{2/3}}{2}.
\eea
$c_{fx}$ are equal theoretically to one and are fitted in practice.
%%%
In Eq. (\ref{massform}), $m_1$, $m_2$, $c_x$ and $\sigma$ are universal for the heavy-heavy systems. $c_{0x}$ vary with different {\rts}.

For the heavy-light systems ($m_1\to\infty$ and $m_2\to0$), Eq. (\ref{qsse}) simplifies to
\begin{eqnarray}\label{qssenr}
M\Psi_{d,t}({\bfr})=\left[m_1+{\bfpa}+V_{d,t}\right]\Psi_{d,t}({\bfr}).
\end{eqnarray}
By employing the Bohr-Sommerfeld quantization approach \cite{Brau:2000st} and using Eq. (\ref{qssenr}), the parameterized formula can be written as \cite{Chen:2022flh,Chen:2021kfw}
\bea\label{rtmeson}
M=m_R+\beta_x\sqrt{x+c_{0x}}\;(x=l,\,n_r,\,L,\,N_r).
\eea
$\beta_x$ is in Eq. (\ref{parabm}), and
with
\bea\label{cxcons}
c_{c}=\sqrt{\sigma'},\quad c_{l,L}=2,\quad c_{n_r,N_r}=\sqrt{2\pi}.
\eea
For the heavy-light systems, the common choice of $m_R$ is \cite{Selem:2006nd,Chen:2021kfw,Jakhad:2023mni,Chen:2014nyo,Chen:2017fcs,Veseli:1996gy,Jia:2018vwl}
\bea\label{mrm1}
m_R=m_1.
\eea
%%%

The usual {\rt}, Eq. (\ref{rtmeson}) with (\ref{mrm1}), is obtained in the limit $m_1\to\infty$ and $m_2\to0$.
There are different ways of including the light constituent's mass \cite{Selem:2006nd,Nielsen:2018uyn,Sonnenschein:2018fph,MartinContreras:2020cyg,
Chen:2023cws,Chen:2022flh,Chen:2023ngj,Chen:2014nyo,Afonin:2014nya,Sergeenko:1994ck}. Two modified formulas are proposed in Ref. \cite{Chen:2023cws}, which can universally describe both the heavy-light mesons and the heavy-light diquarks. One is
Eq. (\ref{rtmeson}) with $m_R$ in (\ref{parabm}), where $m_2$ is the light constituent's mass. Another reads
\bea\label{mrtf}
M=m_R+\sqrt{\beta_x^2(x+c_{0x})+\kappa_{x}m^{3/2}_2(x+c_{0x})^{1/4}}
\eea
if $m_2{\ll}M$, where
\bea\label{mrfp}
m_R=m_1+C',\quad \kappa_x=\frac{4}{3}\sqrt{{\pi}\beta_x},
\eea
where $\beta_x$ is in (\ref{parabm}).
Equation (\ref{rtmeson}) with (\ref{parabm}) is an extension of \cite{Afonin:2014nya}
\bea\label{afoninequ}
M=m_1+m_2+\sqrt{a(n_r+{\alpha}l+b)}
\eea
and the formula \cite{Chen:2022flh}
\bea\label{rmfnpb}
(M-m_1-m_2-C)^2=\alpha_x(x+c_0)^{\gamma}
\eea
while (\ref{mrtf}) with (\ref{mrfp}) is based on the results in \cite{Selem:2006nd,Sonnenschein:2018fph}.
As $m_2=0$, these two modified formulas, Eq. (\ref{rtmeson}) with (\ref{parabm}) and Eq. (\ref{mrtf}) with (\ref{mrfp}), become identical. As $m_2=0$ and $C$ is neglected, these two modified formulas reduce to the usual {\rt} formula for the heavy-light systems, i.e., (\ref{rtmeson}) with (\ref{mrm1}).
Although they give different behavior of $m_2$, Eq. (\ref{rtmeson}) with (\ref{parabm}) and Eq. (\ref{mrtf}) with (\ref{mrfp}) produce consistent results for $l,\,n_r<10$ and have the same behavior $M{\sim}x^{1/2}$ \cite{Chen:2023cws}.

For the light systems ($m_1,\,m_2\to0$), Eq. (\ref{qsse}) simplifies to
\begin{eqnarray}\label{qsseur}
M\Psi_{d,t}({\bfr})=\left[2{\bfpa}+V_{d,t}\right]\Psi_{d,t}({\bfr}).
\end{eqnarray}
By employing the Bohr-Sommerfeld quantization approach \cite{Brau:2000st}, the parametrized formula can be written as \cite{Chen:2022flh,Chen:2021kfw}
\bea\label{rtmesonur}
M=\beta_x\sqrt{x+c_{0x}}\;(x=l,\,n_r,\,L,\,N_r).
\eea
For the light systems, the parameters read as
\bea\label{massformhl}
c_{c}=\sqrt{\sigma'},\; c_{l,L}=2\sqrt{2},\; c_{n_r,N_r}=2\sqrt{\pi}.
\eea
%%%
According to the results in Refs.  \cite{Selem:2006nd,Sonnenschein:2018fph}, similar to Eq. (\ref{mrtf}) \cite{Chen:2023cws}, we suggest
\bea\label{lightm3}
M=C'+\sqrt{\beta_x^2(x+c_{0x})+\kappa_{x}\left(m_1^{3/2}+m_2^{3/2}\right)
(x+c_{0x})^{1/4}}.
\eea
Based on Eq. (\ref{afoninequ}) \cite{Afonin:2014nya} and (\ref{rmfnpb}), we obtain a modified {\rt} relation for the light-light systems in Ref. \cite{Chen:2023ngj}
\bea\label{lighlight}
M=m_R+\beta_x\sqrt{x+c_{0x}},
\eea
where
\bea
c_c=\sqrt{\sigma'},\; c_{l,L}= 2\sqrt{2},\; c_{n_r,N_r}=2\sqrt{\pi}.
\eea
$m_R$ is in Eq. (\ref{parabm}) with $m_{1,2}$ being the masses of the light constituents.

\begin{table}[!phtb]
\caption{The coefficients for the heavy-heavy systems (HHS), the heavy-light systems (HLS), and the light-light systems (LLS).}  \label{tab:eparam}
\centering
\begin{tabular*}{0.47\textwidth}{@{\extracolsep{\fill}}cccc@{}}
\hline\hline
                   & HHS &  HLS & LLS  \\
\hline
$\nu$    & $2/3$ & $1/2$  & $1/2$  \\
$c_c$    & $\left({\sigma'^2}/{\mu}\right)^{1/3}$    & $\sqrt{\sigma'}$ & $\sqrt{\sigma'}$   \\
$c_{l,\,L}$    & $3/2$ & $2$ & $2\sqrt{2}$  \\
$c_{n_r,\,N_r}$ & ${\left(3\pi\right)^{2/3}}/{2}$      & $\sqrt{2\pi}$   & $2\sqrt{\pi}$ \\
\hline
\hline
\end{tabular*}
\end{table}

When Eq. (\ref{rtmeson}) with (\ref{parabm}) is employed to discuss the heavy-light systems, and Eq. (\ref{lighlight}) with (\ref{parabm}) is used to discuss the light systems, summarizing Eqs. (\ref{massform}), (\ref{rtmeson}) with (\ref{parabm}) and (\ref{lighlight}) with (\ref{parabm}), we have a general form of the {\rts} \cite{Chen:2022flh,plan}
\begin{align}\label{massfinal}
M=&m_R+\beta_x(x+c_{0x})^{\nu}\,\,(x=l,\,n_r,\,L,\,N_r),\nonumber\\
m_R=&m_1+m_2+C',\quad \beta_x=c_{fx}c_xc_{c},
\end{align}
where ${\nu}$, $c_x$ and $c_{c}$ are listed in Table \ref{tab:eparam}. $c_{fx}$ are theoretically equal to one and are fitted in practice. $c_{0x}$ vary with different {\rts}. Eq. (\ref{massfinal}) can be employed to discuss various systems including the heavy-heavy systems, the heavy-light systems, and the light-light systems: diquarks, mesons, baryons, and tetraquarks \cite{Chen:2023djq,Chen:2023ngj,Chen:2023web}.

It should be noticed that the general form (\ref{massfinal}) is provisional. Because there are different methods to include the masses of the light constituents. To distinguish the better one, more theoretical and experimental data are needed. In addition to this, the parameter values are universal for both the heavy-heavy systems and the heavy-light systems \cite{Chen:2023cws,Feng:2023txx}, while the parameter values should change for the light systems \cite{Chen:2023ngj} to obtain agreeable results.

\begin{table*}[!phtb]
\caption{Behaviors of three series of {\rt} for different tetraquarks in the diquark picture. $Q=c,\,b$ and $q=u,\,d,\,s$. $M$ is the mass of the tetraquark. $x_{\lambda}=L,\,N_{r}$ are the angular momentum quantum number and the radial quantum number for the $\lambda$ mode. $x_{\rho_1}=l_{1},\,n_{r_{1}}$ are for the $\rho_1$ mode and $x_{\rho_2}=l_{2},\,n_{r_2}$ are for the $\rho_2$ mode. ${\bqm}$ denotes the indefinite result. }\label{tab:rtbehav}
\centering
\begin{tabular*}{1.0\textwidth}{@{\extracolsep{\fill}}cccccccc@{}}
\hline\hline
    &       &   \multicolumn{3}{c}{Trajectory behavior $M{\sim}x^{\nu}$} &   \multicolumn{3}{c}{Trajectory behavior $M^2{\sim}x^{\nu'}$}    \\
  &    & $\rho_1$-mode &  $\rho_2$-mode & $\lambda$-mode  & $\rho_1$-mode &  $\rho_2$-mode & $\lambda$-mode  \\
\hline
1 & $(q_1q'_1)(\bar{q}_2\bar{q}'_2)$ &
$x_{\rho_1}^{1/2}{\bqm}$, $x_{\rho_1}^{3/4}{\bqm}$ &
$x_{\rho_2}^{1/2}{\bqm}$, $x_{\rho_2}^{3/4}{\bqm}$ &
$x_{\lambda}^{1/2}{\bqm}$,  $x_{\lambda}^{2/3}{\bqm}$ & $x_{\rho_1}^{1/2}{\bqm}$, $x_{\rho_1}^{3/4}{\bqm}$  &
$x_{\rho_2}^{1/2}{\bqm}$, $x_{\rho_2}^{3/4}{\bqm}$  &
$x_{\lambda}{\bqm}$,  $x_{\lambda}^{2/3}{\bqm}$ \\
2&  $(q_1q'_1)(\bar{Q}_2\bar{q}_2)$     &
 $x_{\rho_1}^{1/2}{\bqm}$, $x_{\rho_1}^{3/4}{\bqm}$ &
 $x_{\rho_2}^{1/2}$ &
 $x_{\lambda}^{1/2}{\bqm}$,  $x_{\lambda}^{2/3}{\bqm}$ &
 $x_{\rho_1}^{1/2}{\bqm}$, $x_{\rho_1}^{3/4}{\bqm}$ &
 $x_{\rho_2}^{1/2}$ &
  $x_{\lambda}^{1/2}{\bqm}$,  $x_{\lambda}^{2/3}{\bqm}$   \\
3& $(q_1Q_1)(\bar{q}_2\bar{q}'_2)$     &
  $x_{\rho_1}^{1/2}$ &
  $x_{\rho_2}^{1/2}{\bqm}$, $x_{\rho_2}^{3/4}{\bqm}$ &
   $x_{\lambda}^{1/2}{\bqm}$,  $x_{\lambda}^{2/3}{\bqm}$ &
  $x_{\rho_1}^{1/2}$ &
  $x_{\rho_2}^{1/2}{\bqm}$, $x_{\rho_2}^{3/4}{\bqm}$ &
  $x_{\lambda}^{1/2}{\bqm}$,  $x_{\lambda}^{2/3}{\bqm}$ \\
4& $(q_1q'_1)(\bar{Q}_2\bar{Q}'_2)$     &
 $x_{\rho_1}^{1/2}{\bqm}$, $x_{\rho_1}^{3/4}{\bqm}$ &
 $x_{\rho_2}^{2/3}$ &
 $x_{\lambda}^{1/2}{\bqm}$,  $x_{\lambda}^{2/3}{\bqm}$ &
 $x_{\rho_1}^{1/2}{\bqm}$, $x_{\rho_1}^{3/4}{\bqm}$ &
 $x_{\rho_2}^{2/3}$ &
  $x_{\lambda}^{1/2}{\bqm}$,  $x_{\lambda}^{2/3}{\bqm}$   \\
5& $(Q_1Q'_1)(\bar{q}_2\bar{q}'_2)$     &
  $x_{\rho_1}^{2/3}$ &
  $x_{\rho_2}^{1/2}{\bqm}$, $x_{\rho_2}^{3/4}{\bqm}$ &
   $x_{\lambda}^{1/2}{\bqm}$,  $x_{\lambda}^{2/3}{\bqm}$ &
  $x_{\rho_1}^{2/3}$ &
  $x_{\rho_2}^{1/2}{\bqm}$, $x_{\rho_2}^{3/4}{\bqm}$ &
  $x_{\lambda}^{1/2}{\bqm}$,  $x_{\lambda}^{2/3}{\bqm}$ \\
6& $(q_1Q_1)(\bar{Q}_2\bar{q}_2)$     &
  $x_{\rho_1}^{1/2}$ &
  $x_{\rho_2}^{1/2}$ &
  $x_{\lambda}^{2/3}$ &
  $x_{\rho_1}^{1/2}$ &
  $x_{\rho_2}^{1/2}$ &
  $x_{\lambda}^{2/3}$ \\
7& $(q_1Q_1)(\bar{Q}_2\bar{Q}'_2)$     &
  $x_{\rho_1}^{1/2}$ &
  $x_{\rho_2}^{2/3}$ &
  $x_{\lambda}^{2/3}$ &
  $x_{\rho_1}^{1/2}$ &
  $x_{\rho_2}^{2/3}$ &
  $x_{\lambda}^{2/3}$ \\
8& $(Q_1Q'_1)(\bar{Q}_2\bar{q}_2)$     &
  $x_{\rho_1}^{2/3}$ &
  $x_{\rho_2}^{1/2}$ &
  $x_{\lambda}^{2/3}$ &
  $x_{\rho_1}^{2/3}$ &
  $x_{\rho_2}^{1/2}$ &
  $x_{\lambda}^{2/3}$     \\
9& $(Q_1Q'_1)(\bar{Q}_2\bar{Q}'_2)$     &
  $x_{\rho_1}^{2/3}$ &
  $x_{\rho_2}^{2/3}$ &
  $x_{\lambda}^{2/3}$ &
  $x_{\rho_1}^{2/3}$ &
  $x_{\rho_2}^{2/3}$ &
  $x_{\lambda}^{2/3}$     \\
\hline
\hline
\end{tabular*}
\end{table*}

\subsection{Behaviors of the {\rts} for various tetraquarks}\label{sec:rtbh}

The {\rts} behave differently in different energy regions \cite{Chen:2022flh,Chen:2021kfw}. The {\rt} behaviors for baryons are discussed in Ref. \cite{plan}.
In this subsection, we discuss the {\rt} behaviors for various tetraquarks.

Taken as an example, tetraquarks $(cc)(\bar{q}\bar{q}')$ are discussed,
where $q$ and $q'$ are the light quarks, $u$, $d$, or $s$.
The diquark $(cc)$ is the doubly heavy diquark, therefore, the diquark {\rts} are clear, see Eq. (\ref{massform}),
\bea\label{rhoa}
M_{\rho_1}{\sim}x^{2/3}_{\rho_1}\,\;(x_{\rho_1}=l_1,\,n_{r_1}),
\eea
where $M_{\rho_1}$ is mass of the diquark $(cc)$. The antidiquark $(\bar{q}\bar{q}')$ is the light diquark. Although they have indefinite forms [see Eqs. (\ref{lightm3}) and (\ref{lighlight}) with (\ref{parabm})], the light-diquark {\rts} have the same behavior,
\bea\label{rhob}
M_{\rho_2}{\sim}x^{1/2}_{\rho_2}\,\;(x_{\rho_2}=l_2,\,n_{r_2}),
\eea
where $M_{\rho_2}$ is mass of the antidiquark $(\bar{q}\bar{q}')$.
Generally speaking, the diquark {\rts} are not the same as the $\rho$-trajectories of tetraquarks.

When discussing the $\lambda$-mode, the {\rt} relation for the tetraquarks $(cc)(\bar{q}\bar{q}')$ is not clear. We will be confronted with two problems. One is when the light antidiquark can be regarded as being heavy, which will affect not only the behaviors of the ${\lambda}$-{\trs} but also the behaviors of the $\rho$-{\trs}. Another is how to introduce the mass of the light constituent if the antidiquark is light, which will affect the behavior of the $\rho$-{\trs}.
If the light antidiquark $(\bar{q}\bar{q}')$ is regarded as being light, the tetraquarks $(cc)(\bar{q}\bar{q}')$ are the heavy-light systems, therefore,  $\lambda$-trajectories behave as
\bea
M{\sim}x^{1/2}_{\lambda}\;(x_{\lambda}=L,\,N_r)
\eea
no matter whether Eq. (\ref{rtmeson}) with (\ref{parabm}) or Eq. (\ref{mrtf}) with (\ref{mrfp}) is chosen as the $\lambda$-trajectory formula.
When Eq. (\ref{mrtf}) is chosen as the $\lambda$-trajectory relation, it will give $M{\sim}M_{\rho_1},\,M_{\rho_2}^{3/2}$ while Eq. (\ref{rtmeson}) with (\ref{parabm}) gives $M{\sim}M_{\rho_1},\,M_{\rho_2}$ due to different ways to include the light constituent's mass $M_{\rho_2}$ [corresponding to $m_{2}$ in Eq. (\ref{rtmeson}) with (\ref{parabm}) and Eq. (\ref{mrtf}) with (\ref{mrfp})]. They give two different behaviors of the $\rho$-trajectories. Using Eqs. (\ref{rhoa}) and (\ref{rhob}),  Eq. (\ref{rtmeson}) with (\ref{parabm}) gives $M{\sim}x_{\rho_1}^{2/3},\,x_{\rho_2}^{3/4}$ while Eq. (\ref{mrtf}) with (\ref{mrfp}) gives $M{\sim}x_{\rho_1}^{2/3},\,x_{\rho_2}^{1/2}$.
%%%
[When discussing the $\rho$-trajectories, highly excited antidiquarks are involved. The first radially and orbitally excited $(ud)$ are about $1.3$ {\gev}. The masses of the $1^1p_1$ states of $(us)$ and $(ss)$ are much heavier, $1.46$ {\gev} and $1.65$ {\gev} \cite{Chen:2023ngj}, respectively. They approximate or even are larger than the mass of the charm quark. Therefore, the light antidiquark can be regarded as being heavy.]
If the light antidiquark is regarded as being heavy, $\lambda$-trajectories behave as
\bea\label{lamrtb}
M{\sim}x_{\lambda}^{2/3}
\eea
according to Eq. (\ref{massform}) with (\ref{parabm}). Using Eqs. (\ref{massform}), (\ref{parabm}), (\ref{rhoa}), and (\ref{rhob}), we have
$M{\sim}M_{\rho_1},\,M_{\rho_2}{\sim}x_{\rho_1}^{2/3},\,x_{\rho_2}^{1/2}$.

Other types of tetraquarks can be discussed by similar way. The {\rt} behaviors of different types of tetraquarks are listed in Table \ref{tab:rtbehav}.

When the diquark (antidiquark) is doubly heavy or heavy-light, it is clear how to introduce its large mass, see Eqs. (\ref{massform}) with (\ref{parabm}) and Eq. (\ref{rtmeson}) with (\ref{parabm}) or Eq. (\ref{mrtf}) with (\ref{mrfp}). After choosing the {\rt} formulas for the heavy-light diquarks, for example, (\ref{rtmeson}) with (\ref{parabm}), the {\rt} formulas can be written explicitly, and then the behaviors of the $\lambda$-trajectory and the $\rho$-trajectory are definite. Using Eq. (\ref{massfinal}), the {\rt} relations for the tetraquarks $(Q_1q_1)(\bar{Q}_2\bar{q}_2)$ read
\begin{align}\label{t2q}
M&=m_{R{\lambda}}+\beta_{x_{\lambda}}(x_{\lambda}+c_{0x_{\lambda}})^{2/3}\;(x_{\lambda}=L,\,N_r),\nonumber\\
M_{\rho_1}&=m_{R\rho_1}+\beta_{x_{\rho_1}}\sqrt{x_{\rho_1}+c_{0x_{\rho_1}}}\;(x_{\rho_1}=l_{1},\,n_{r_1}),\nonumber\\
M_{\rho_2}&=m_{R\rho_2}+\beta_{x_{\rho_2}}\sqrt{x_{\rho_2}+c_{0x_{\rho_2}}}\;(x_{\rho_2}=l_{2},\,n_{r_2}),
\end{align}
where
\begin{align}\label{pa2qQ}
m_{R{\lambda}}&=M_{\rho_1}+M_{\rho_2}+C,\nonumber\\
m_{R\rho_1}&=m_{Q_1}+m_{q_1}+C/2,\nonumber\\
m_{R\rho_2}&=m_{Q_2}+m_{q_2}+C/2,\nonumber\\
\beta_{L}&=\frac{3}{2}\left(\frac{\sigma^2}{\mu_{\lambda}}\right)^{1/3}c_{fL},\nonumber \\
%\end{align}
%\begin{align}%\label{pa2qQ}
\beta_{N_r}&=\frac{(3\pi)^{2/3}}{2}\left(\frac{\sigma^2}{\mu_{\lambda}}\right)^{1/3}c_{fN_r},\nonumber\\
\mu_{\lambda}&=\frac{M_{\rho_1}M_{\rho_2}}{M_{\rho_1}+M_{\rho_2}},\nonumber\\
\beta_{l_1}&=\sqrt{2\sigma}c_{fl_1},\quad \beta_{n_{r_1}}=\sqrt{\pi\sigma}c_{fn_{r_1}},\nonumber\\
\beta_{l_2}&=\sqrt{2\sigma}c_{fl_2},\quad \beta_{n_{r_2}}=\sqrt{\pi\sigma}c_{fn_{r_2}}.
\end{align}
The {\rt} relations for the tetraquarks $(Q_1Q'_1)(\bar{Q}_2\bar{q}_2)$ read
\begin{align}\label{t3q}
M&=m_{R{\lambda}}+\beta_{x_{\lambda}}(x_{\lambda}+c_{0x_{\lambda}})^{2/3}\;(x_{\lambda}=L,\,N_r),\nonumber\\
M_{\rho_1}&=m_{R\rho_1}+\beta_{x_{\rho_1}}(x_{\rho_1}+c_{0x_{\rho_1}})^{2/3}\;(x_{\rho_1}=l_{1},\,n_{r_1}),\nonumber\\
M_{\rho_2}&=m_{R\rho_2}+\beta_{x_{\rho_2}}\sqrt{x_{\rho_2}+c_{0x_{\rho_2}}}\;(x_{\rho_2}=l_{2},\,n_{r_2}),
\end{align}
where
\begin{align}\label{pa3qQ}
m_{R{\lambda}}&=M_{\rho_1}+M_{\rho_2}+C,\nonumber\\
m_{R\rho_1}&=m_{Q_1}+m_{Q'_1}+C/2,\nonumber\\
m_{R\rho_2}&=m_{Q_2}+m_{q_2}+C/2,\nonumber\\
\beta_{L}&=\frac{3}{2}\left(\frac{\sigma^2}{\mu_{\lambda}}\right)^{1/3}c_{fL},\; \beta_{N_r}=\frac{(3\pi)^{2/3}}{2}\left(\frac{\sigma^2}{\mu_{\lambda}}\right)^{1/3}c_{fN_r},\nonumber\\
\mu_{\lambda}&=\frac{M_{\rho_1}M_{\rho_2}}{M_{\rho_1}+M_{\rho_2}},\;
\mu_{\rho_1}=\frac{m_{Q_1}m_{Q'_1}}{m_{Q_1}+m_{Q'_1}},\nonumber\\
\beta_{l_1}&=\frac{3}{2}\left(\frac{\sigma^2}{4\mu_{\rho_1}}\right)^{1/3}c_{fl_1},\; \beta_{l_2}=\sqrt{2\sigma}c_{fl_2},\nonumber\\ \beta_{n_{r_1}}&=\frac{(3\pi)^{2/3}}{2}\left(\frac{\sigma^2}{4\mu_{\rho_1}}\right)^{1/3}c_{fn_{r_1}},
\quad \beta_{n_{r_2}}=\sqrt{\pi\sigma}c_{fn_{r_2}}.
\end{align}
The tetraquark {\rt} relations for $(Q_1q_1)(\bar{Q}_2\bar{Q}'_2)$ can be obtained easily by the similar way as obtaining the {\rt} formulas for  $(Q_1Q'_1)(\bar{Q}_2\bar{q}_2)$.
The {\rt} relations for the tetraquarks $(Q_1Q'_1)(\bar{Q}_2\bar{Q}'_2)$ read
\begin{align}\label{t4q}
M&=m_{R{\lambda}}+\beta_{x_{\lambda}}(x_{\lambda}+c_{0x_{\lambda}})^{2/3}\;(x_{\lambda}=L,\,N_r),\nonumber\\
M_{\rho_1}&=m_{R\rho_1}+\beta_{x_{\rho_1}}(x_{\rho_1}+c_{0x_{\rho_1}})^{2/3}\;(x_{\rho_1}=l_{1},\,n_{r_1}),\nonumber\\
M_{\rho_2}&=m_{R\rho_2}+\beta_{x_{\rho_2}}(x_{\rho_1}+c_{0x_{\rho_1}})^{2/3}\;(x_{\rho_2}=l_{2},\,n_{r_2}),
\end{align}
where
\begin{align}\label{pa4qQ}
m_{R{\lambda}}&=M_{\rho_1}+M_{\rho_2}+C,\nonumber\\
m_{R\rho_1}&=m_{Q_1}+m_{Q'_1}+C/2,\nonumber\\
m_{R\rho_2}&=m_{Q_2}+m_{Q'_2}+C/2,\nonumber\\
\beta_{L}&=\frac{3}{2}\left(\frac{\sigma^2}{\mu_{\lambda}}\right)^{1/3}c_{fL},\; \beta_{N_r}=\frac{(3\pi)^{2/3}}{2}\left(\frac{\sigma^2}{\mu_{\lambda}}\right)^{1/3}c_{fN_r},\nonumber\\
\mu_{\lambda}&=\frac{M_{\rho_1}M_{\rho_2}}{M_{\rho_1}+M_{\rho_2}},\;
\mu_{\rho_1}=\frac{m_{Q_1}m_{Q'_1}}{m_{Q_1}+m_{Q'_1}},\nonumber\\
\mu_{\rho_2}&=\frac{m_{Q_2}m_{Q'_2}}{m_{Q_2}+m_{Q'_2}},\;
\beta_{l_1}=\frac{3}{2}\left(\frac{\sigma^2}{4\mu_{\rho_1}}\right)^{1/3}c_{fl_1},\; \nonumber\\ \beta_{n_{r_1}}&=\frac{(3\pi)^{2/3}}{2}\left(\frac{\sigma^2}{4\mu_{\rho_1}}\right)^{1/3}c_{fn_{r_1}},\nonumber\\
\beta_{l_2}&=\frac{3}{2}\left(\frac{\sigma^2}{4\mu_{\rho_2}}\right)^{1/3}c_{fl_2},\nonumber\\ \beta_{n_{r_2}}&=\frac{(3\pi)^{2/3}}{2}\left(\frac{\sigma^2}{4\mu_{\rho_2}}\right)^{1/3}c_{fn_{r_2}}.
\end{align}
Due to ambiguity, the tetraquark {\rt} formulas for the tetraquarks containing the light diquark or the light antidiquark are not given.

\subsection{Discussions}\label{subsec:rtdisc}

The form of the {\rts} depends not only on energy regions but also on the confining potential \cite{Chen:2022flh,Chen:2021kfw}. In this work, the employed confining potential is linear.
Moreover, various mixings are not considered when discussing the behaviors of the tetraquark {\rts}, see Table \ref{tab:rtbehav}. When the mixings are considered, the behaviors of the tetraquark {\rts} will become complex.

In potential models, the curvature of the {\rts} is related to the dynamic equation and the confining potential \cite{Chen:2018bbr}.
In Ref. \cite{Chen:2023web}, we suggest that the {\rts} for the diquarks, mesons, baryons and tetraquarks are concave downward in the $(M^2,\,x)$ planes. In fact, the baryon trajectories and the tetraquark {\rts} discussed in Refs. \cite{Chen:2023web,Chen:2023djq} actually are the $\lambda$-trajectories.
We can see from Table \ref{tab:rtbehav} that both the $\lambda$-trajectories and the $\rho$-trajectories for various tetraquarks are all concave downward in the $(M^2,\,x)$ plane. For the light systems, the trajectories approximate linear as the masses are neglected and become concave when the masses are considered.

If the diquark quark contents have the same flavors as the antidiquark contents, there are degeneracy in masses. The $\rho_1$-mode excited states have the same masses as the $\rho_2$-mode excited states.

\section{Determination of $c_{fx_{\lambda}}$ and $c_{0x_{\lambda}}$ for the $\lambda$-modes}\label{sec:appcfx}

In this section, we determine the values of $c_{fx_{\lambda}}$ and $c_{0x_{\lambda}}$ for the $\lambda$-modes of the {\twh}.
Apply Eq. (\ref{massfinal}) to fit the {\rts} for the doubly heavy mesons, the $\lambda$-modes of heavy-heavy baryons composed of one heavy quark and one doubly heavy (or one heavy-light) diquark and the $\lambda$-modes of heavy-heavy tetraquarks composed of one doubly heavy (or heavy-light) diquark and one doubly heavy (or heavy-light) antidiquark.
The quality of a fit is measured by the quantity $\chi^2$ defined by
\bea
\chi^2=\frac{1}{N-1}\sum^{N}_{i=1}\left(\frac{M_{fi}-M_{ei}}{M_{ei}}\right)^2,
\eea
where $N$ is the number of points on the trajectory, $M_{fi}$ is fitted value and $M_{ei}$ is the experimental value or the theoretical value of the $i$-th particle mass. The parameters are determined by minimizing $\chi^2$.

\begin{table}[!phtb]
\caption{The spin averaged masses of the radially excited states of mesons and baryons, and the masses of tetraquarks (in ${\gev}$).}  \label{tab:cfr}
\centering
\begin{tabular*}{0.5\textwidth}{@{\extracolsep{\fill}}ccccccc@{}}
\hline\hline
           & $1S$  &  $2S$  &  $3S$  & $4S$  &  $5S$  &  $6S$  \\
\hline
$c\bar{c}$  & $3.0687$  & $3.6740$ & $4.0273$ & $4.4115$ &$4.8305$ & $5.1640$ \\
$c\bar{b}$  & 6.3184   & 6.8793   &  7.2500  &  7.6030  &  7.9430  &   \\
$b\bar{b}$  &9.4450    & 10.0173   & 10.3486   & 10.5778   & 10.8645   & 11.0812  \\
$\Omega_{ccb}$ &7.9940    &8.4097  &    &    &    &   \\
$\Omega_{cbb}$  & 11.2107   &11.6967    &    &    &    &   \\
$[cs][\bar{c}\bar{s}]$
                & 4.051   & 4.604   &    &    &    &   \\
$[bs][\bar{b}\bar{s}]$
                & 10.662   &11.111    &    &    &    &   \\
\hline\hline
\end{tabular*}
\end{table}

\begin{table}[!phtb]
\caption{The spin averaged masses of the orbitally excited states of mesons and baryons, and the masses of tetraquarks (in ${\gev}$).}  \label{tab:cfo}
\centering
\begin{tabular*}{0.5\textwidth}{@{\extracolsep{\fill}}ccccccc@{}}
\hline\hline
            & $1S$  &  $1P$  &  $1D$  & $1F$  &  $1G$  &  $1H$  \\
\hline
$c\bar{c}$ & 3.0687   & 3.5253   & 3.8186   & 4.0720   & 4.3435   & 4.5931  \\
$c\bar{b}$ & 6.3184   & 6.7486   & 7.0261   & 7.2720   & 7.4879   &   \\
$b\bar{b}$ & 9.4450   & 9.8997   & 10.1629   & 10.3467   & 10.5127   & 10.6711  \\
$\Omega_{ccb}$  &7.9940 & 8.2643   & 8.4741   &    &    &   \\
$\Omega_{cbb}$  &11.2107    &11.5261    &11.8053    &    &    &   \\
$[cs][\bar{c}\bar{s}]$
              & 4.051   &4.466    &4.728    &    &    &   \\
$[bs][\bar{b}\bar{s}]$
                &10.662    &11.002    &11.216    &    &    &   \\
\hline\hline
\end{tabular*}
\end{table}

The masses used are listed in Tables \ref{tab:cfr} and \ref{tab:cfo}. For the mesons and baryons, the listed data are the spin averaged masses. If the corresponding states are experimentally determined, the PDG data from Ref. \cite{pdg:2024} are used. For the undetermined states, the theoretical data from Ref. \cite{Ebert:2011jc,Faustov:2021qqf} are used. (In Ref. \cite{Ebert:2011jc}, the mass of the $\Omega_{cbb}$ $(7/2)^{+}$(1D) state is not provided, so we use the value $11.807$ ${\gev}$.) The theoretical data for the tetraquarks are from \cite{Faustov:2021hjs,Faustov:2022mvs}.
Beside the bound states' masses, some parameters are listed in Table \ref{tab:parmv}. The masses of the axial vector diquarks are $m_{\{cc\}}=3.14$ ${\gev}$, $m_{\{bb\}}=9.63$ ${\gev}$ \cite{Feng:2023txx}. The masses of the scalar diquarks $[cs]$ and $[bs]$ are calculated by using the parameters in Table \ref{tab:parmv} and Eq. (\ref{massfinal}) or (\ref{t2q}) \cite{Chen:2023cws}.

\begin{figure*}[!phtb]
\centering
\subfigure[]{\label{subfigure:fiterr}\includegraphics[scale=0.45]{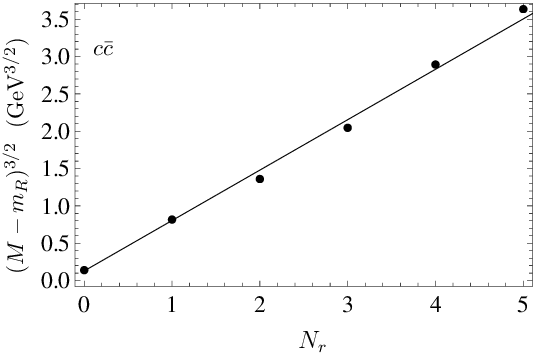}}
\subfigure[]{\label{subfigure:fiterr}\includegraphics[scale=0.45]{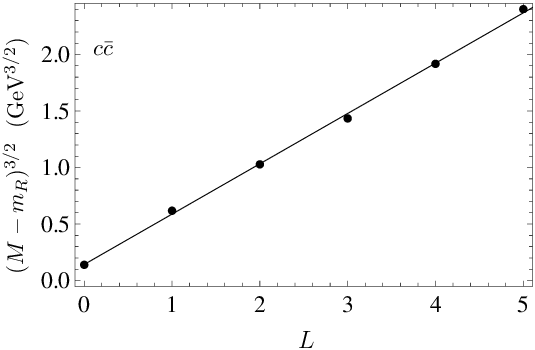}}
\subfigure[]{\label{subfigure:fiterr}\includegraphics[scale=0.45]{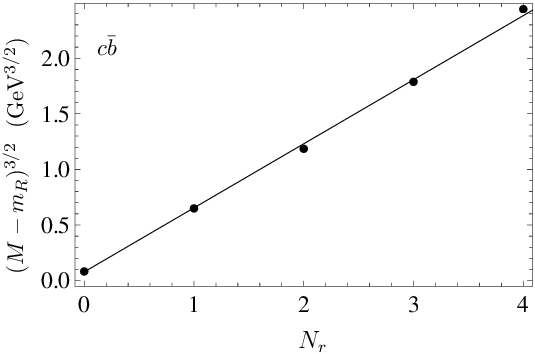}}
\subfigure[]{\label{subfigure:fiterr}\includegraphics[scale=0.45]{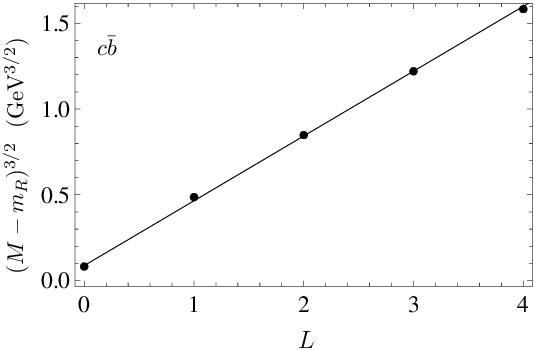}}
\subfigure[]{\label{subfigure:fiterr}\includegraphics[scale=0.45]{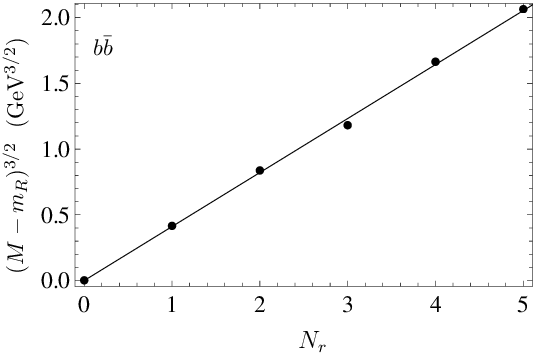}}
\subfigure[]{\label{subfigure:fiterr}\includegraphics[scale=0.45]{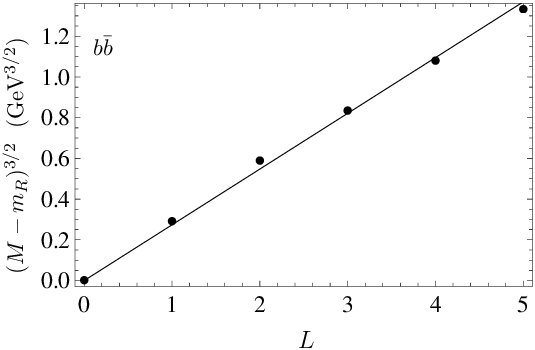}}
\subfigure[]{\label{subfigure:fiterr}\includegraphics[scale=0.45]{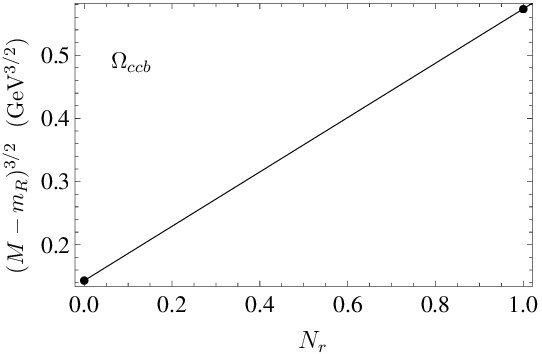}}
\subfigure[]{\label{subfigure:fiterr}\includegraphics[scale=0.45]{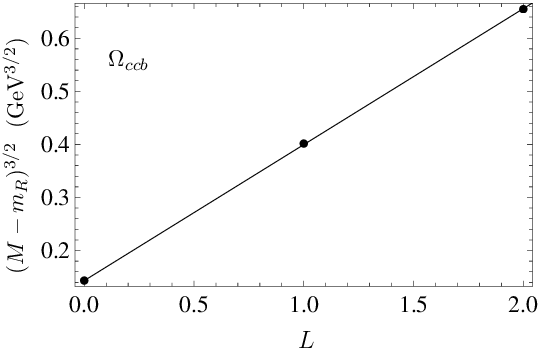}}
\subfigure[]{\label{subfigure:fiterr}\includegraphics[scale=0.45]{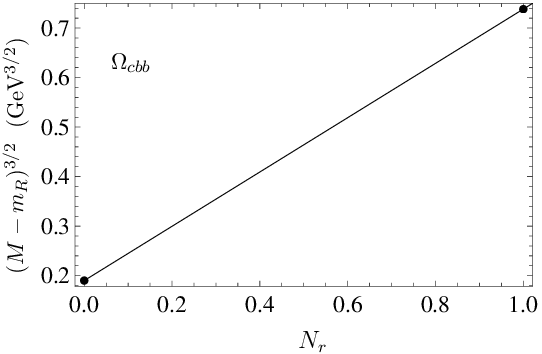}}
\subfigure[]{\label{subfigure:fiterr}\includegraphics[scale=0.45]{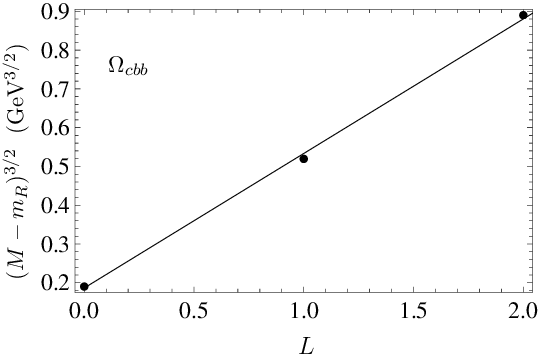}}
\subfigure[]{\label{subfigure:fiterr}\includegraphics[scale=0.45]{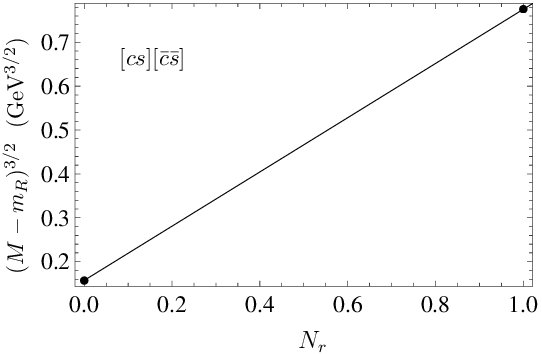}}
\subfigure[]{\label{subfigure:fiterr}\includegraphics[scale=0.45]{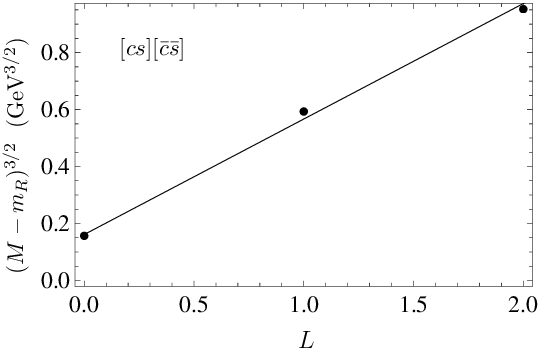}}
\subfigure[]{\label{subfigure:fiterr}\includegraphics[scale=0.45]{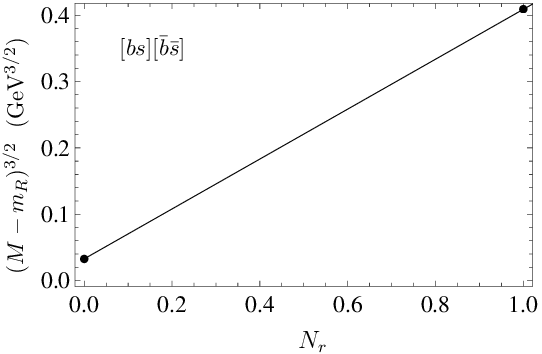}}
\subfigure[]{\label{subfigure:fiterr}\includegraphics[scale=0.45]{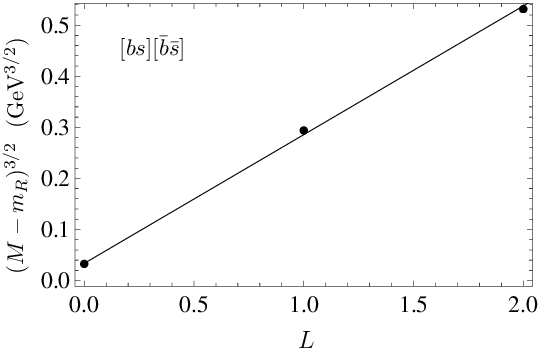}}
\caption{The radial and orbital {\rts} (the black lines) for mesons, baryons and tetraquarks. The dots are the spin averaged masses of mesons and baryons and the masses of tetraquarks. $N_r$ and $L$ are the radial and orbital quantum numbers for the $\lambda$-mode, respectively. The used data are listed in Tables \ref{tab:cfr}-\ref{tab:fp}.}\label{fig:rgapp}
\end{figure*}

The fitted parameters are listed in Table \ref{tab:fp}. The fitted {\rts} are plotted in the $((M-m_R)^{3/2},\;x)$ plane, see Fig. \ref{fig:rgapp}. The results are consistent with those in Ref. \cite{Chen:2023djq}.

\begin{figure}[!phtb]
\centering
\subfigure[]{\label{subfigure:cfa}\includegraphics[scale=0.7]{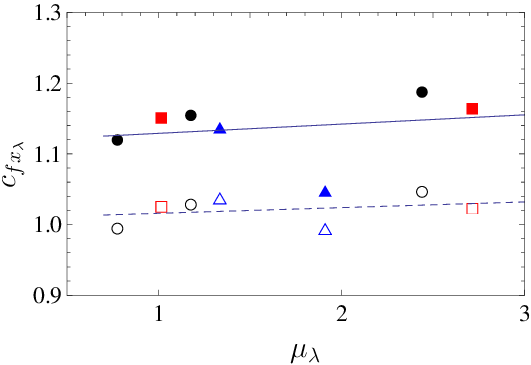}}
\subfigure[]{\label{subfigure:cfb}\includegraphics[scale=0.7]{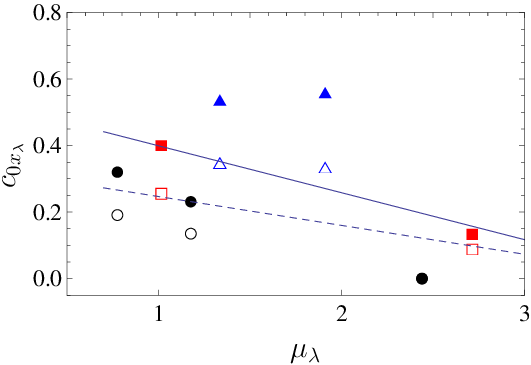}}
\caption{The fitted $c_{fx_{\lambda}}$ [Fig. \ref{subfigure:cfa}] and $c_{0x_{\lambda}}$ [Fig. \ref{subfigure:cfb}], ($x_{\lambda}=L,\,N_r$). The data are listed in Table \ref{tab:fp}. $\mu_{\lambda}$ is the reduced masses, see Eq. (\ref{rdmu}) or (\ref{pa2qQ}). The black filled circles (mesons), the red filled squares (tetraquarks), and the green filled triangles (baryons) are the fitted $c_{fL}$ or $c_{0L}$. The black empty circles (mesons), the red empty squares (tetraquarks), and the green empty triangles (baryons) are the fitted $c_{fN_r}$ or $c_{0N_r}$. The black lines ($c_{fL}$ and $c_{0L}$) and the dashed lines ($c_{fN_r}$ and $c_{0N_r}$) are the linear fit, see Eqs. (\ref{fitcfxl}) and (\ref{fitcfxnr}). }\label{fig:rgc0fx}
\end{figure}

The fitted $c_{fx_{\lambda}}$ and $c_{0x_{\lambda}}$ vary across different {\rts} for different states and for different systems, see Fig. \ref{fig:rgc0fx} and Table \ref{tab:fp}. In case of the baryons and tetraquarks, these parameters are sensitive to diquark masses. For example, $c_{fN_r}=1.218$ when $[bc]=6.38$ ${\gev}$ \cite{Feng:2023txx} while $c_{fN_r}=1.027$ when $[bc]=6.519$ ${\gev}$, as determined by fitting the radial {\rt} for the tetraquark $[bc][\bar{b}\bar{c}]$ \cite{Faustov:2022mvs}.

According to Eq. (\ref{t2q}), $c_{fx_{\lambda}}$ and $c_{0x_{\lambda}}$ are needed to determine a {\rt} as $m_{R_{\lambda}}$ can be calculated by using Eq. (\ref{pa2qQ}) and parameters in Table \ref{tab:parmv}. Two or more states on the {\rt} are needed to obtain $c_{fx_{\lambda}}$ and $c_{0x_{\lambda}}$. We choose $c_{fx_{\lambda}}$ and $c_{0x_{\lambda}}$ by fitting these parameters from other systems. By fitting the data in Table \ref{tab:fp}, we have
\begin{eqnarray}
c_{fL}=&1.116 + 0.013\mu_{\lambda},\; c_{0L}=0.540- 0.141\mu_{\lambda}, \label{fitcfxl}\\
c_{fN_r}=&1.008 + 0.008\mu_{\lambda}, \;  c_{0N_r}=0.334 - 0.087\mu_{\lambda},\label{fitcfxnr}
\end{eqnarray}
where $\mu_{\lambda}$ is the reduced masses, see Eq. (\ref{rdmu}) or (\ref{pa2qQ}). Eqs. (\ref{fitcfxl}) and (\ref{fitcfxnr}) are obtained when $\mu_{\lambda}<3.83$ ${\gev}$. And the formulas for the $\mu_{\lambda}>3.83$ ${\gev}$ are not considered here. As more experimental data or more theoretical data become available, the fitted formulas will be refined.

\begin{table*}[!phtb]
\caption{The fitted $c_{fx_{\lambda}}$ and $c_{0x_{\lambda}}$.}  \label{tab:fp}
\centering
\begin{tabular*}{1\textwidth}{@{\extracolsep{\fill}}cccccccc@{}}
\hline\hline
            & $c\bar{c}$  &  $c\bar{b}$  &  $b\bar{b}$  & $\Omega_{ccb}$  &  $\Omega_{cbb}$  &  $[cs][\bar{c}\bar{s}]$ & $[bs][\bar{b}\bar{s}]$  \\
\hline
 $(c_{fN_r},\,c_{0N_r})$   &(0.994,\,0.191)  &(1.028,\,0.135)  &(1.046,\,0.0)  &(0.994,\,0.333)  & (1.037,\,0.347) &(1.025,\,0.255)  &(1.022,\,0.087)  \\
 $(c_{fL},\,c_{0L})$       &(1.120,\,0.320)  & (1.154,\,0.231) & (1.187,\,0.0) &(1.047,\,0.559)  &(1.137,\,0.537)  &(1.151,\,0.399)  &(1.164,\,0.133)  \\
\hline\hline
\end{tabular*}
\end{table*}

\end{document}